\documentclass[12pt]{article}

\usepackage[margin=1in]{geometry}
\usepackage{amsmath,amssymb,amsthm,bm}
\usepackage{graphicx}
\usepackage{booktabs}
\usepackage{setspace}
\usepackage{url}
\usepackage{placeins}
\usepackage[colorlinks=true,citecolor=blue,linkcolor=blue,urlcolor=blue]{hyperref}
\usepackage{lineno}
\usepackage{multirow}
\usepackage{makecell}
\graphicspath{{Figs/}}
\usepackage[american]{babel}
\usepackage{csquotes}

\usepackage[
  backend=biber,
  style=authoryear,
  natbib=true,
  giveninits=true,
  uniquename=false,
  maxcitenames=2,
  maxbibnames=99,
  doi=true,
  url=false,
  isbn=false,
  eprint=true
]{biblatex}

\theoremstyle{definition}
\newtheorem{definition}{Definition}
\newtheorem{assumption}{Assumption}
\newtheorem{lemma}{Lemma}
\newtheorem{theorem}{Theorem}

\newtheorem{proposition}{Proposition}
\newtheorem{remark}{Remark}

\newcommand{\LEAF}{\operatorname{LEAF}}
\newcommand{\E}{\operatorname{E}}
\newcommand{\Var}{\operatorname{Var}}
\newcommand{\Cov}{\operatorname{Cov}}
\newcommand{\med}{\operatorname{median}}

\newcommand{\bone}{\bm{1}}
\newcommand{\bj}{\bm{j}}
\newcommand{\bY}{\bm{Y}}
\newcommand{\bmu}{\bm{\mu}}
\newcommand{\bSigma}{\bm{\Sigma}}
\newcommand{\beps}{\bm{\varepsilon}}
\newcommand{\br}{\bm{r}}
\newcommand{\HCAL}{\widehat H_{\mathrm{CAL}}}

\title{LEAF: Calibrated and Robust Wavelet Estimation of the Hurst Exponent}
\author{Vijini Lakmini and Brani Vidakovic\\Department of Statistics, Texas A\&M University}
\date{\today}

\begin{document}
%\linenumbers
\maketitle

\begin{abstract}
Wavelet logscale regression for the Hurst exponent $H$ of fractional Brownian
motion (fBm) is sensitive to outliers and biased in finite samples. We
introduce the Log-Energy with Adaptive Floor (LEAF) ordinate, which averages
the logarithms of squared wavelet coefficients after adding a floor
proportional to a pilot estimate of the level variance. If standardized
coefficients share a common law across levels, a floor proportional to the
true level variance preserves the population slope, and a consistent pilot
preserves it asymptotically. We derive the asymptotic variance and influence
function, and identify the variance-minimizing floor for median-pilot LEAF,
whose influence grows only logarithmically. With outliers of ten increment
standard deviations in up to $1\%$ of observations, median-pilot LEAF reduces root mean squared error by a factor of $1.8$ to $2.6$ relative
to calibrated log-energy. At $H=0.5$ with $1\%$ outliers, the error falls
from $0.139$ to $0.059$.

For small to moderate $H$, the main source of bias is the departure of
expected wavelet energies from the ideal power law at fine scales. We
estimate $H$ from an estimating equation calibrated by simulation under fBm.
Calibrating the mean logscale diagram alone delivers nearly all of the gain:
it nearly eliminates bias in matched simulations and improves on the
Abry--Veitch estimator most at small $H$. Simulations also show how
scale-dependent tails can erode the gains of a robust pilot and how
misspecification can undermine calibration.
\end{abstract}


%======================================================================
\section{Introduction}
%======================================================================

Fractional Brownian motion (fBm) is one of the most widely used models for self-similar processes. Its law is governed by the Hurst exponent $H\in(0,1)$, which determines both the roughness of the paths and the dependence of the increments. Accurate estimation of $H$ matters in hydrology, telecommunications, finance, geophysics, image analysis, and biomedical signal processing, and a large literature is devoted to it; see \citet{Zhang2024Hurst} and \citet{Vimalajeewa2026WaveletHurst} for recent overviews and \citet{Ju2023Fractal,Fujimoto2024Scaling,Averty2025Hurst} for recent non-wavelet methods.

Wavelet methods are standard because the variance of the wavelet coefficients of fBm follows a power law in scale
\citep{GacheFlandrinGarreau1991,Flandrin1992,AbryVeitch1998,VeitchAbry1999,PercivalWalden2000}. The classical logscale estimator averages the squared coefficients at each level, takes logarithms, and regresses them on the level index. Because the expected logarithm of an average is not the logarithm of its expectation, \citet{AbryVeitch1998} introduced an analytical correction, giving the widely used Abry--Veitch (AV) estimator with digamma bias correction and trigamma weights. Semiparametric alternatives, such as the wavelet Whittle estimator, have been developed and compared with Fourier methods by \citet{MoulinesRoueffTaqqu2008} and \citet{FayEtAl2009}. Two practical difficulties remain, and this paper addresses them separately.

The first is finite-sample bias that is not logarithmic. For sampled fBm and a finite discrete wavelet transform, the true variances $\sigma_j^2=\E d_{j,k}^2$ at the fitted levels do not lie on the line $\log_2\sigma_j^2=a+(2H+1)j$. Possible sources are the initialization of the multiresolution analysis with sampled data \citep{VeitchTaqquAbry2000} and the boundary coefficients. The discrepancy is a property of the population, so no correction of the logarithm removes it, and \citet{Wu2020} reports that an alternative initialization does not remove the empirical bias either. In our experiments the discrepancy at small $H$ comes almost entirely from the fine scales and persists when boundary-affected coefficients are removed; boundary coefficients matter mainly for $H\ge0.8$ (Section~\ref{sec:sources}). For $N=1024$, the Daubechies wavelet with two vanishing moments, and levels $2$ to $7$, the expected level energies imply, with the AV weights, exponents of approximately $0.099$ when $H=0.2$ and $0.468$ when $H=0.5$ (Table~\ref{tab:decomp}). This structural component dominates the bias of the weighted least squares log-energy slope for $H\le0.6$ (Section~\ref{sec:sources}). The logarithmic effect is nearly constant, about $-0.011$ to $-0.014$ in $H$, of which the AV correction removes $0.009$. A natural remedy is to calibrate the estimator by simulation from the model. Monte Carlo corrections and confidence intervals have been tabulated for classical estimators of long-range dependence \citep{Weron2002}. More generally, matching an auxiliary statistic with its simulated expectation is the principle of indirect inference \citep{GourierouxMonfortRenault1993,Smith1993} and of simulation-based bias correction \citep{MacKinnonSmith1998,GourierouxPhillipsYu2010,GuerrierEtAl2019}. For wavelet variances, the generalized method of wavelet moments \citep{GuerrierEtAl2013} and its robust version \citep{GuerrierEtAl2022} match empirical and model-implied wavelet variances. These ideas have not, to our knowledge, been used to calibrate the full finite-sample mean and covariance of the logscale diagram for the estimation of $H$.

The second difficulty is sensitivity to outliers and to departures from Gaussianity. Ordinates based on the mean energy have an influence that grows quadratically with the size of a coefficient. A large literature replaces the mean energy by robust alternatives. Log-first estimators, which average logarithms of individual coefficient energies, appeared for self-similar $\alpha$-stable processes, where second moments may not exist \citep{AbryTaqquPesquet2000,StoevPipirasTaqqu2002}. \citet{SoltaniSimardBoichu2004} combined neighboring coefficients, and \citet{Park2009Robust} studied the average-log ordinate from a robustness perspective. \citet{MondalPercival2012} proposed M-estimation of the wavelet variance on the scale of the log-squared coefficients, with a bias correction from the moments of $\log\chi^2_1$ \citep{BartlettKendall1946}. \citet{KangVidakovic2017} proposed the MEDL and MEDLA estimators based on medians of log-squared nondecimated coefficients, and \citet{FengVidakovic2017} proposed trimean estimators; these were applied in image analysis by \citet{FengMeiVidakovic2018}. Robust scale estimators, such as the $Q_n$ estimator of \citet{RousseeuwCroux1993}, were studied for wavelet log-regression and under long-range dependence by \citet{KouamoLevyLeducMoulines2013} and \citet{LevyLeducEtAl2011}. Other robust estimators use quantiles of discrete variations \citep{Coeurjolly2008,AchardCoeurjolly2010}, robust wavelet regressions \citep{ShenZhuLee2007}, or robust combinations of pairs of levels \citep{HamiltonEtAl2022}. Coefficientwise logarithms bound the influence of large energies, but energies near zero make them highly variable: under the Gaussian model the asymptotic variance of the average-log ordinate is $2.47$ times that of the log-energy ordinate.

We stabilize the logarithm with a floor proportional to the level variance. The Log-Energy with Adaptive Floor (LEAF) ordinate is
$$
Y_j(\lambda)=\frac1{n_j}\sum_{k=1}^{n_j}\log_2\left(d_{j,k}^2+\lambda\widehat\sigma_j^2\right),
$$
where $\widehat\sigma_j^2$ is a pilot estimate of the level variance, such as the mean or the median energy, and $\lambda\ge0$ is dimensionless. The ordinate interpolates between the average-log ordinate ($\lambda=0$) and the pilot ordinate $\log_2\widehat\sigma_j^2$ ($\lambda\to\infty$). A floor of this kind has a precedent in quasi-maximum likelihood estimation of stochastic volatility models, where the logarithm of squared returns is unstable for returns near zero. For this problem \citet[Section~9.3]{Fuller1996} proposed the modified transformation $\log(y^2+c\,s^2)-c\,s^2/(y^2+c\,s^2)$, with $s^2$ the sample mean of $y^2$ and $c=0.02$, and \citet{BreidtCarriquiry1996} used it to improve quasi-maximum likelihood estimation. Without its correction term, this is the LEAF transform with a mean pilot and $\lambda=0.02$. Shifted logarithms with offsets tied to the scale of the data are also standard for variance stabilization in genomics \citep{DurbinEtAl2002,AhlmannEltzeHuber2023}. We apply the floor level by level to wavelet coefficients, allow a robust pilot, and calibrate the resulting logscale diagram against its finite-sample expectation under sampled fBm. Comparing ordinates under the same calibration shows where each gain arises: calibrating the mean accounts for nearly all of the improvement on clean fBm, while median-pilot LEAF gives the lowest errors in our experiments with large additive outliers in up to $1\%$ of the observations.

\subsection{Contributions}

The paper makes four contributions. The first is a calibrated estimating equation for $H$. For any vector of logscale ordinates $\bY$, whether log-energy, AV, LEAF or log-median, we define the estimator $\HCAL$ as the root of
$$
q_N(h)\{\bY-\bmu_N(h)\}=0,
$$
where $\bmu_N(h)$ and $\bSigma_N(h)$ are the finite-sample mean and covariance of $\bY$ under fBm with exponent $h$, estimated once from an independent simulation bank, and $q_N(h)$ is the generalized least squares (GLS) slope functional built from $\bSigma_N(h)$. In matched fBm simulations the estimator substantially reduces the bias, and it requires neither tuning nor a preliminary estimate. Replacing $q_N(h)$ by fixed AV weights gives a mean-only calibrated version, which needs only the simulated mean, delivers almost all of the improvement, and is the simplest version to use in practice. Under uniform identification and local regularity conditions the root exists, is unique near $H$, and has asymptotic variance equal to the GLS variance divided by $s_N(H)^2$, where $s_N(H)=\frac12q_N(H)\bmu_N'(H)$ measures how fast the expected ordinate vector moves with $H$ (Theorem~\ref{thm:root}). The one-step update $\widehat H_1+\frac12q_N(\widehat H_1)\{\bY-\bmu_N(\widehat H_1)\}$ from a preliminary estimator with the usual rate retains the error $\{1-s_N(H)\}(\widehat H_1-H)$; at small $H$, where $s_N$ exceeds $2$, it overshoots and its iterates diverge, so the root must be computed (Section~\ref{sec:cal}).

The second contribution is a theory of LEAF ordinates (Section~\ref{sec:leaf-theory}). We prove that a floor proportional to the true level variance is exactly slope-neutral under Gaussianity and that a floor proportional to a consistent pilot is asymptotically slope-neutral, we show that a common standardized coefficient law across levels suffices for slope neutrality and that departures from it generally bias the slope, and we derive the asymptotic variance and the influence function of the ordinate for any asymptotically linear pilot. For a single level with independent Gaussian coefficients no ordinate has smaller asymptotic variance than log-energy, and with the mean pilot LEAF differs from log-energy only by terms of order $\lambda^{-2}$, for any data, once $\lambda$ exceeds the largest normalized coefficient energy. With the median pilot the variance has an interior minimum at $\lambda^\ast\approx0.096$, where the efficiency relative to log-energy is $0.60$, compared with $0.41$ for the average-log ordinate and $0.37$ for the log-median ordinate. The influence of a single coefficient grows quadratically, with coefficient $w(\lambda)<1$, for the mean pilot; for the median pilot it is unbounded but grows only logarithmically, and it is bounded at the log-median endpoint.

The third contribution is a simulation study showing that, under Gaussian fBm, calibration rather than the choice of ordinate explains the gains (Section~\ref{sec:gauss}). With exact Davies--Harte fBm, $N\in\{512,1024,4096\}$, and $M=5000$ common paths, the calibrated estimator has residual bias of at most $0.003$. For $N=1024$, with both estimators restricted to the calibration range $[0.01,0.99]$, it reduces the mean squared error of AV by a factor from $1.3$ to $23$, and without the restriction the factor reaches $97$ at $H=0.1$, where AV is negative on average. Paired experiments show that calibrating only the mean, with fixed AV weights, achieves nearly all of the gain, that full covariance weighting adds $2$--$4\%$, that removing boundary-affected coefficients does not change the picture, and that omitting the finest fitted level reduces the bias of AV at the price of a larger variance (Section~\ref{sec:sources}). Calibrated mean-pilot LEAF and calibrated log-energy have similar accuracy, with paired differences in mean squared error below $2\%$ for $\lambda\ge1$, and the empirical efficiencies of all ordinates follow the asymptotic theory. The exact Gaussian likelihood is $2.6$ to $5.1$ times more accurate than the calibrated wavelet estimators, which indicates what is lost by the logscale summary in this idealized setting.

The fourth contribution concerns robustness (Section~\ref{sec:robust}). With additive outliers of size ten increment standard deviations in up to $1\%$ of the observations, calibrated LEAF with the median pilot and $\lambda=0.1$ is the most accurate estimator considered; at $H=0.5$ with $1\%$ outliers, for example, it lowers the root mean squared error from $0.139$ for calibrated log-energy to $0.059$. Its root mean squared error is $1.8$ to $2.6$ times smaller than that of calibrated log-energy and smaller than those of the average-log and log-median ordinates, which are its endpoints, while the Gaussian maximum likelihood estimator, the most accurate method on clean data, deteriorates sharply under these outliers, especially at larger $H$. With smaller outliers, of size five, a mean pilot with a small floor is competitive, and the mean pilot with $\lambda=0.5$, which costs about one percent in root mean squared error under the Gaussian model, improves accuracy under heavy-tailed increments for $H\ge0.5$. We also report where robust ordinates and calibration fail: heavy tails whose shape changes across scales bias median-pilot ordinates, a calibration bank from the wrong model introduces bias, and nonlinear transforms of antipersistent fGn and jumps in the increments change the low-frequency scaling, so that the latent fBm exponent becomes the target of a misspecified analysis.

The paper is organized as follows. Section~\ref{sec:setup} introduces the setting and the LEAF ordinate, and Section~\ref{sec:leaf-theory} develops the properties of LEAF ordinates. Section~\ref{sec:cal} develops the calibrated estimator. Sections~\ref{sec:gauss} and~\ref{sec:robust} report simulation studies under Gaussian fBm and under contamination and non-Gaussianity. Section~\ref{sec:disc} concludes with recommendations. Proofs and implementation details are in the Appendix.

%======================================================================
\section{Setting and the LEAF ordinate}
\label{sec:setup}
%======================================================================

This section fixes the notation. We first recall wavelet logscale regression for fBm, together with the classical log-energy ordinate and the Abry--Veitch estimator, and then define the LEAF transform, the LEAF ordinate, and the two pilot estimators of the level variance used throughout the paper.

\subsection{Wavelet logscale regression}

Let $X$ be fBm with Hurst exponent $H$ observed at $N=2^J$ equispaced times, and let $d_{j,k}$, $k=1,\dots,n_j$, be its discrete wavelet coefficients at level $j$, where $j=1$ is the finest level and resolution decreases as $j$ increases. The coefficients are Gaussian with mean zero. For a wavelet with at least one vanishing moment, the coefficients whose support does not reach the boundary extension are stationary within a level, and, in the continuous-time idealization,
$$
\sigma_j^2=\E d_{j,k}^2=K_\psi(H)\,2^{(2H+1)j}.
$$
Coefficients affected by the boundary extension, a few at each level for half-point symmetric extension, have different variances, so within-level stationarity is an idealization for them. Logscale regression fits a line to ordinates $Y_j$ over a set of levels $\mathcal J=\{j_1,\dots,j_K\}$, and estimates $H=(\beta-1)/2$ from the fitted slope $\beta$. The classical ordinate is the log-energy
$$
Y_j^{\mathrm{LE}}=\log_2\widehat\sigma_{j,\mathrm{mean}}^2,
\qquad
\widehat\sigma_{j,\mathrm{mean}}^2=\frac1{n_j}\sum_{k=1}^{n_j}d_{j,k}^2 .
$$
Under an independent Gaussian idealization $n_j\widehat\sigma_{j,\mathrm{mean}}^2/\sigma_j^2\sim\chi^2_{n_j}$, so
$$
\E Y_j^{\mathrm{LE}}=\log_2\sigma_j^2+g(n_j),
\qquad
g(n)=\frac{\psi(n/2)}{\log 2}-\log_2\frac n2,
\qquad
\Var Y_j^{\mathrm{LE}}=\frac{\psi'(n_j/2)}{(\log2)^2}.
$$
The AV estimator regresses $Y_j^{\mathrm{LE}}-g(n_j)$ on $j$ by weighted least squares (WLS) with weights $(\log 2)^2/\psi'(n_j/2)$ \citep{AbryVeitch1998,VeitchAbry1999}.

\subsection{The LEAF transform and ordinate}

\begin{definition}[LEAF ordinate]
For a floor $C>0$ define $\LEAF_C(d)=\log_2(d^2+C)$. For a pilot estimate $\widehat\sigma_j^2$ of the level variance and $\lambda\ge0$, the LEAF ordinate at level $j$ is
$$
Y_j(\lambda)=\frac1{n_j}\sum_{k=1}^{n_j}\log_2\left(d_{j,k}^2+\lambda\widehat\sigma_j^2\right),
$$
with the limits $Y_j(0)=n_j^{-1}\sum_k\log_2 d_{j,k}^2$ (average-log) and, after removal of the constant $\log_2\lambda$, $Y_j(\infty)=\log_2\widehat\sigma_j^2$ (pilot ordinate).
\end{definition}

We consider two pilots, the mean energy $\widehat\sigma_{j,\mathrm{mean}}^2$ and the median energy
$$
\widehat\sigma_{j,\mathrm{med}}^2=\frac{\med_k(d_{j,k}^2)}{m_1},
\qquad
m_1=\med(\chi^2_1)\approx0.4549364 .
$$
With the mean pilot, $Y_j(\infty)$ is the log-energy ordinate; with the median pilot, it is the log-median ordinate. Trimmed and winsorized means are possible intermediate pilots; the theory below covers any pilot with an asymptotically linear representation, and we do not pursue them in the simulations.

%======================================================================
\section{Properties of LEAF ordinates}
\label{sec:leaf-theory}
%======================================================================

This section studies the LEAF ordinate as a statistic of the coefficients at a single level. We first show that a proportional floor leaves the population slope unchanged, then expand the ordinate around its limit when the pilot is estimated, and use the expansion to obtain its asymptotic variance, its efficiency relative to log-energy, its influence function, and its consistency. Throughout the section $d_{j,k}=\sigma_jZ_{j,k}$ with $Z_{j,k}\sim N(0,1)$. This is exact for the interior coefficients of fBm and an idealization for boundary-affected coefficients. Within-level dependence does not affect the marginal calculations; asymptotic variances are stated for independent coefficients, and dependence enters through effective sample sizes, which the calibration of Section~\ref{sec:cal} handles exactly.

\subsection{Slope neutrality and scale equivariance}

\begin{definition}[Gaussian calibration function]
For $\rho\ge0$ let $b(\rho)=\E\log_2(Z^2+\rho)$, $Z\sim N(0,1)$.
\end{definition}

\begin{lemma}[Gaussian LEAF decomposition]
\label{lem:decomp}
For a deterministic floor $C_j>0$,
$$
\E\LEAF_{C_j}(d_{j,k})=\log_2\sigma_j^2+b(\rho_j),
\qquad
\rho_j=C_j/\sigma_j^2 .
$$
\end{lemma}

\begin{proof}
Factor $\sigma_j^2$ from $\sigma_j^2Z_{j,k}^2+C_j$ and apply the definition of $b$.
\end{proof}

\begin{proposition}[Fixed and proportional population floors]
\label{prop:neutral}
If $C_j=C$ is constant across levels, then $\rho_j=C/\{K_\psi(H)2^{(2H+1)j}\}$ varies with $j$ and the floor biases the slope. If $C_j=\lambda\sigma_j^2$, then $\E\LEAF_{C_j}(d_{j,k})=\log_2\sigma_j^2+b(\lambda)$ at every level, so the floor shifts the intercept and leaves the slope unchanged.
\end{proposition}

\begin{lemma}[Scale equivariance]
\label{lem:scale}
If every coefficient is multiplied by $a\ne0$ and the pilot is scale-equivariant, $\widehat\sigma_j^2\mapsto a^2\widehat\sigma_j^2$, then $Y_j(\lambda)\mapsto Y_j(\lambda)+2\log_2|a|$ for every $\lambda$ and every level.
\end{lemma}

\begin{proof}
$\log_2(a^2d_{j,k}^2+\lambda a^2\widehat\sigma_j^2)=2\log_2|a|+\log_2(d_{j,k}^2+\lambda\widehat\sigma_j^2)$.
\end{proof}

Proposition~\ref{prop:neutral} is an exact statement about a floor proportional to the true level variance. With an estimated pilot, the floor $\lambda\widehat\sigma_j^2$ is random and correlated with the coefficients, and the finite-sample shift $\E Y_j(\lambda)-\log_2\sigma_j^2$ depends on $n_j$ and on the dependence within level $j$. It therefore differs across levels; slope neutrality holds only asymptotically (Theorem~\ref{thm:cons}). These finite-sample differences are part of the distortion that the calibration of Section~\ref{sec:cal} absorbs.

Lemma~\ref{lem:scale} implies that the slope of any LEAF ordinate vector, and its finite-sample distribution after removal of a common shift, do not depend on the amplitude of the process. Calibration can therefore be performed with unit-scale fBm.

\begin{remark}[Role of a common standardized law]
\label{rem:shape}
Proposition~\ref{prop:neutral} uses only that the standardized coefficients $d_{j,k}/\sigma_j$ have the same distribution at all levels. If the standardized law $G_j$ depends on $j$, the shift becomes $b_j(\lambda)=\E_{G_j}\log_2(U^2+\lambda\kappa_j)$, with $\kappa_j$ the ratio of the pilot's limit to $\sigma_j^2$, which in general varies with $j$ and biases the slope. This happens for non-Gaussian processes whose fine-scale coefficients are heavy-tailed while coarse-scale coefficients, being sums of many increments, are nearly Gaussian. The log-energy ordinate is the least affected because $\E d_{j,k}^2=\sigma_j^2$ holds for every law; ordinates with small floors or median pilots are the most affected. Section~\ref{sec:robust} shows this effect.
\end{remark}

\subsection{First-order expansion with an estimated pilot}

Write $\widehat\sigma_j^2=\sigma_j^2(1+\eta_j)$ and let $Y_j^0(\lambda)$ be the ordinate computed with the true $\sigma_j^2$. Since $d_{j,k}^2+\lambda\sigma_j^2\ge\lambda\sigma_j^2$,
$$
\left|Y_j(\lambda)-Y_j^0(\lambda)\right|
=\left|\frac1{n_j}\sum_{k=1}^{n_j}\log_2\left\{1+\frac{\lambda\sigma_j^2\eta_j}{d_{j,k}^2+\lambda\sigma_j^2}\right\}\right|
\le-\log_2(1-|\eta_j|),
\qquad |\eta_j|<1,
$$
so the error of the pilot propagates only through its relative error. A Taylor expansion gives, to first order,
$$
Y_j(\lambda)-Y_j^0(\lambda)\approx\frac{w(\lambda)}{\log2}\,\eta_j,
\qquad
w(\lambda)=\lambda\E\left\{(Z^2+\lambda)^{-1}\right\}\in[0,1].
$$
The weight $w$ increases from $w(0)=0$ to $w(\infty)=1$. Since the regression slope annihilates constants, a common relative pilot error $\eta_j\equiv\eta$ affects only the intercept, to first order in the population expansion.

Suppose the pilot is asymptotically linear,
$$
\eta_j=\frac1{n_j}\sum_{k=1}^{n_j}\phi(Z_{j,k})+o_p(n_j^{-1/2}),
\qquad\E\phi(Z)=0,
$$
with $\phi(z)=z^2-1$ for the mean pilot and
$$
\phi(z)=\frac{1/2-\mathbb 1\{z^2\le m_1\}}{m_1f_1(m_1)}
$$
for the median pilot, where $f_1$ is the $\chi^2_1$ density. Combining the two expansions,
$$
Y_j(\lambda)=\log_2\sigma_j^2+b(\lambda)+\frac1{n_j}\sum_{k=1}^{n_j}\mathrm{IF}_\lambda(Z_{j,k})+o_p(n_j^{-1/2}),
$$
$$
\mathrm{IF}_\lambda(z)=\frac{1}{\log2}\left[\log(z^2+\lambda)-\E\log(Z^2+\lambda)+w(\lambda)\phi(z)\right].
$$
The function $\mathrm{IF}_\lambda$ is the influence function \citep{Hampel1974} of the ordinate as a functional of the standardized coefficient distribution.

\subsection{Asymptotic variance and efficiency}

\begin{theorem}[Asymptotic normality of the LEAF ordinate]
\label{thm:var}
Let the coefficients at level $j$ be independent $N(0,\sigma_j^2)$, let $\lambda>0$ be fixed, and let the pilot be the mean or the median energy. Then, as $n_j\to\infty$,
$$
n_j^{1/2}\left\{Y_j(\lambda)-\log_2\sigma_j^2-b(\lambda)\right\}\Rightarrow N\left(0,V(\lambda)\right),
\qquad
V(\lambda)=\E\,\mathrm{IF}_\lambda(Z)^2 .
$$
We call $V(\lambda)$ the asymptotic variance of the ordinate; the theorem concerns the limiting distribution and does not assert convergence of $n_j\Var Y_j(\lambda)$.
For the mean pilot, $V(\infty)=2/(\log2)^2\approx4.1627$ and $V(0)=\Var\log_2Z^2=\pi^2/\{2(\log 2)^2\}\approx10.2711$.
\end{theorem}

\begin{proposition}[Efficiency bound]
\label{prop:cr}
Under independent $N(0,\sigma_j^2)$ coefficients, every ordinate that is regular under local changes in scale and has the form $\log_2\sigma_j^2+\text{const}+n_j^{-1}\sum_k\mathrm{IF}(Z_{j,k})+o_p(n_j^{-1/2})$ satisfies $\E\,\mathrm{IF}(Z)^2\ge2/(\log 2)^2$. The bound is attained by the log-energy ordinate. Hence $V(\lambda)\ge V(\infty)$ for every pilot and every $\lambda$.
\end{proposition}

Proposition~\ref{prop:cr} concerns a single level with independent Gaussian coefficients. It does not establish optimality of log-energy for the correlated coefficients of fBm, nor for the calibrated estimator of $H$ that combines several levels. Indeed, Section~\ref{sec:gauss} finds mean-pilot LEAF slightly more accurate than log-energy at $H=0.9$. The proof (Appendix~\ref{app:proofs}) is the Cram\'er--Rao bound for $\log\sigma^2$ in the model $N(0,\sigma^2)$, whose Fisher information is $1/2$. Table~\ref{tab:vlam} gives $V(\lambda)$, the efficiency $V(\infty)/V(\lambda)$ relative to log-energy, and the weight $w(\lambda)$. For the mean pilot, $V$ decreases monotonically to the bound; at $\lambda=2$ the excess is $0.44\%$ and at $\lambda=0.5$ it is $4.3\%$. For the median pilot, $V$ has an interior minimum at $\lambda^\ast\approx0.096$, where the efficiency is $0.597$, compared with $0.405$ for the average-log ordinate ($\lambda=0$) and $0.368$ for the log-median ordinate ($\lambda=\infty$). The floor is therefore not merely a numerical safeguard: with a robust pilot it raises the efficiency of coefficientwise logarithms by half.

\begin{table}[ht]
\centering
\begin{tabular}{rcccccc}
\toprule
& \multicolumn{2}{c}{Mean pilot} & \multicolumn{2}{c}{Median pilot} & \\
\cmidrule(lr){2-3}\cmidrule(lr){4-5}
$\lambda$ & $V(\lambda)$ & Efficiency & $V(\lambda)$ & Efficiency & $w(\lambda)$\\
\midrule
0      & 10.271 & 0.405 & 10.271 & 0.405 & 0\\
0.05   & 5.520  & 0.754 & 7.049  & 0.591 & 0.237\\
0.096  & 5.061  & 0.823 & 6.977  & 0.597 & 0.308\\
0.25   & 4.558  & 0.913 & 7.137  & 0.583 & 0.438\\
0.5    & 4.341  & 0.959 & 7.478  & 0.557 & 0.546\\
1      & 4.228  & 0.985 & 8.016  & 0.519 & 0.656\\
2      & 4.181  & 0.996 & 8.710  & 0.478 & 0.758\\
5      & 4.165  & 0.999 & 9.681  & 0.430 & 0.865\\
$\infty$ & 4.163 & 1     & 11.327 & 0.368 & 1\\
\bottomrule
\end{tabular}
\caption{Asymptotic variance $V(\lambda)$ of the LEAF ordinate per coefficient (Theorem~\ref{thm:var}) for independent Gaussian coefficients, efficiency $V_{\mathrm{mean}}(\infty)/V(\lambda)$ relative to log-energy, and the pilot weight $w(\lambda)=\lambda\E(Z^2+\lambda)^{-1}$. Values were computed by numerical quadrature.}
\label{tab:vlam}
\end{table}

The next result makes the equivalence of mean-pilot LEAF and log-energy exact.

\begin{proposition}[Large-floor expansion for the mean pilot]
\label{prop:large}
Let $u_{j,k}=d_{j,k}^2/\widehat\sigma_{j,\mathrm{mean}}^2$ and $m_{j,r}=n_j^{-1}\sum_ku_{j,k}^r$, so that $m_{j,1}=1$. If $\lambda>\max_ku_{j,k}$, then, for any data,
$$
Y_j(\lambda)=Y_j^{\mathrm{LE}}+\log_2\lambda+\frac{1}{\lambda\log 2}
-\frac{m_{j,2}}{2\lambda^2\log 2}
+\sum_{r\ge3}\frac{(-1)^{r+1}m_{j,r}}{r\lambda^r\log2}.
$$
\end{proposition}

\begin{proof}
$Y_j(\lambda)=\log_2(\lambda\widehat\sigma_j^2)+n_j^{-1}\sum_k\log_2(1+u_{j,k}/\lambda)$; expand $\log(1+x)$ for $|x|<1$ and use $m_{j,1}=1$.
\end{proof}

The terms of order $\lambda^0$ and $\lambda^{-1}$ are constant across levels and disappear from any slope or calibrated estimator. Mean-pilot LEAF and log-energy therefore differ by $O_p(\lambda^{-2})$ through the normalized second moments $m_{j,2}$, which explains both the flatness of $V(\lambda)$ for $\lambda\ge1$ and the small paired differences observed in Section~\ref{sec:gauss}.

\subsection{Influence and contamination}

\begin{proposition}[Influence functions]
\label{prop:if}
The influence function of the log-energy ordinate is $(z^2-1)/\log2$. For LEAF with the mean pilot,
$$
\mathrm{IF}_\lambda(z)=\frac1{\log2}\left[\log(z^2+\lambda)-\E\log(Z^2+\lambda)+w(\lambda)(z^2-1)\right],
$$
which grows quadratically with coefficient $w(\lambda)/\log 2<1/\log 2$. For LEAF with the median pilot and finite $\lambda$, $\mathrm{IF}_\lambda(z)=2\log_2|z|+O(1)$ as $|z|\to\infty$, so the influence is unbounded but grows only logarithmically. At the log-median endpoint ($\lambda=\infty$) the influence function is $\phi(z)/\log 2$, which is bounded.
\end{proposition}

Proposition~\ref{prop:if} quantifies the trade-off in Table~\ref{tab:vlam}. With the mean pilot, a smaller floor reduces the quadratic influence by the factor $w(\lambda)$ (for example $w(0.5)=0.55$) at a small cost in variance; with the median pilot, the influence is logarithmic for every finite $\lambda$, and $\lambda^\ast\approx0.1$ gives the most efficient ordinate in this robust class. For gross errors, consider the population mixture in which a fraction $\varepsilon$ of the coefficients at level $j$ have energy $A\sigma_j^2$. The probability limit of the log-energy ordinate shifts by exactly $\log_2(1-\varepsilon+\varepsilon A)$. The same shift holds exactly in a sample in which a fraction $\varepsilon$ of the coefficients is replaced and the retained energies average to $\sigma_j^2$. The shift is large when $\varepsilon A$ is large, whereas to first order the median-pilot LEAF ordinate shifts by $\varepsilon\{\mathrm{IF}_\lambda(\sqrt A)-\E\,\mathrm{IF}_\lambda(Z)\}\approx\varepsilon\log_2 A$. Additive outliers in the time domain produce coefficients with a fixed absolute energy, so $A$ is largest at the fine levels where $\sigma_j^2$ is smallest. The contamination thus bends the logscale diagram and lowers the estimated $H$, and the effect is much weaker for logarithmic influence.

\subsection{Consistency}

\begin{theorem}[Consistency]
\label{thm:cons}
Fix $\mathcal J$ and $\lambda>0$. Suppose that $\log_2\sigma_j^2=a+(2H+1)j$ on $\mathcal J$, that $Y_j^0(\lambda)-\E Y_j^0(\lambda)\to0$ almost surely for each $j\in\mathcal J$, and that $\widehat\sigma_j^2/\sigma_j^2\to1$ almost surely. Then $Y_j(\lambda)-b(\lambda)\to\log_2\sigma_j^2$ almost surely, and the least squares estimator of $H$ based on $Y_j(\lambda)$, with or without the calibration $b(\lambda)$, converges almost surely to $H$.
\end{theorem}

\begin{proof}
By the bound of the previous subsection, $|Y_j(\lambda)-Y_j^0(\lambda)|\le-\log_2(1-|\eta_j|)\to0$. By Lemma~\ref{lem:decomp}, $Y_j^0(\lambda)\to\log_2\sigma_j^2+b(\lambda)$. The least squares slope is a continuous function of the ordinates, and $b(\lambda)$ is common to all levels.
\end{proof}

%======================================================================
\section{Mean- and covariance-calibrated estimation}
\label{sec:cal}
%======================================================================

Theorem~\ref{thm:cons} concerns the limit $N\to\infty$. At fixed $N$, the mean of the ordinate vector departs from the ideal line, and the ordinates are correlated across levels. This section corrects both effects by calibration.

\subsection{Vector formulation}

Let $\bY=(Y_{j_1},\dots,Y_{j_K})^\top$ be any vector of logscale ordinates (log-energy, AV, LEAF, log-median), $\bj=(j_1,\dots,j_K)^\top$, $X=(\bone,\bj)$, and
$$
\bmu_N(H)=\E_H\bY,
\qquad
\bSigma_N(H)=\Cov_H\bY,
$$
where the subscript $N$ indicates that both include all finite-sample effects of the signal length, wavelet, boundary rule, fitted levels, pilot, and floor. For positive definite $\bSigma$, the GLS slope functional is
$$
q_{\bSigma}=e_2^\top(X^\top\bSigma^{-1}X)^{-1}X^\top\bSigma^{-1},
\qquad
q_{\bSigma}\bone=0,
\qquad
q_{\bSigma}\bj=1,
$$
and we write $q_N(H)=q_{\bSigma_N(H)}$. Decompose
$$
\bmu_N(H)=\alpha_N(H)\bone+(2H+1)\bj+\br_N(H),
$$
where $\br_N(H)$ is the finite-sample mean distortion. It contains the structural departure of $\sigma_j^2$ from the power law, the nonlinearity of the ordinate, and the effect of dependence. By Lemma~\ref{lem:scale}, $\bmu_N(H)$ is determined up to a multiple of $\bone$ and $\bSigma_N(H)$ is determined exactly by unit-scale fBm.

\begin{proposition}[AV and log-energy have the same calibrated estimator]
\label{prop:av}
If two ordinate vectors differ by a deterministic vector, $\bY'=\bY-\bm g$, then $\bmu'_N=\bmu_N-\bm g$, $\bSigma'_N=\bSigma_N$, and $\bY'-\bmu_N'=\bY-\bmu_N$. In particular, the AV ordinates $Y_j^{\mathrm{LE}}-g(n_j)$ and the log-energy ordinates give identical calibrated estimators.
\end{proposition}

\subsection{The calibrated estimating equation}

\begin{definition}[Calibrated estimator]
Let
$$
G_N(h;\bY)=q_N(h)\{\bY-\bmu_N(h)\}.
$$
The calibrated estimator $\HCAL$ is the root of $G_N(h;\bY)=0$ in $h\in(0,1)$.
\end{definition}

Replacing $q_N(h)$ by a fixed slope vector $q$ with $q\bone=0$ and $q\bj=1$, such as the AV weights, gives the \emph{mean-only calibrated} estimator, the root of $q\{\bY-\bmu_N(h)\}=0$. It corrects the same mean distortion without using the covariance.

The function $G_N$ compares the observed logscale diagram with the diagram expected under fBm with exponent $h$, using the GLS slope as the metric. This is indirect inference \citep{GourierouxMonfortRenault1993,Smith1993}, with $\bY$ as the auxiliary statistic and $h\mapsto\bmu_N(h)$ as the binding function, and the same principle underlies simulation-based bias correction \citep{MacKinnonSmith1998,GourierouxPhillipsYu2010,GuerrierEtAl2019}. In the wavelet setting it is close to the generalized method of wavelet moments \citep{GuerrierEtAl2013}, which matches empirical wavelet variances to model-implied ones, and to its robust version \citep{GuerrierEtAl2022}. Machine-learning estimators of $H$ trained on simulated paths \citep{Raubitzek2023} perform a similar calibration implicitly. Our construction differs in three ways: the auxiliary statistics are logarithmic ordinates; the binding function is the exact finite-sample mean, obtained by simulation, rather than the population wavelet variance, so boundary and initialization effects are absorbed; and robustness is obtained by the choice of ordinate within the LEAF family.

Define the expected covariance-only estimate and the local sensitivity of the estimating equation,
$$
\overline H_N(h)=\frac{q_N(h)\bmu_N(h)-1}{2},
\qquad
s_N(H)=\frac12q_N(H)\bmu_N'(H)=1+\frac12q_N(H)\br_N'(H).
$$
Under the ideal law $\br_N\equiv0$ we have $s_N\equiv1$. In general $s_N(H)$ measures how fast the expected ordinate vector moves with $H$, as seen through the slope functional fixed at $H$; it is the quantity that enters the expansions below. Because the GLS weights depend on $h$, $s_N$ is not the derivative of $\overline H_N$. Differentiating the identities $q_N(h)\bone=0$ and $q_N(h)\bj=1$ gives $q_N'(h)\bone=q_N'(h)\bj=0$, and hence
$$
\overline H_N'(h)=s_N(h)+\frac12q_N'(h)\bmu_N(h)=s_N(h)+\frac12q_N'(h)\br_N(h),
$$
so the two differ only through the distortion $\br_N$. For $N=1024$ they differ by less than $0.2$ on $[0.05,0.9]$ (Figure~\ref{fig:binding}).

\begin{proposition}[Oracle properties]
\label{prop:oracle}
Let $\widehat H^{\mathrm{or}}=H+\frac12q_N(H)\{\bY-\bmu_N(H)\}$. Then $\E_H\widehat H^{\mathrm{or}}=H$ exactly and
$$
\Var_H\widehat H^{\mathrm{or}}=V_N(H)=\frac14e_2^\top\{X^\top\bSigma_N(H)^{-1}X\}^{-1}e_2,
$$
the smallest variance among linear slope estimators $q\bY$ with $q\bone=0$ and $q\bj=1$. The covariance-only estimator $\{q_N(H)\bY-1\}/2$ has the same variance and bias $c_N(H)=\frac12q_N(H)\br_N(H)$, so its mean squared error exceeds $V_N(H)$ by $c_N(H)^2$.
\end{proposition}

\subsection{One-step updates and why the root is needed}

\begin{figure}[ht]
\centering
\includegraphics[width=0.98\textwidth]{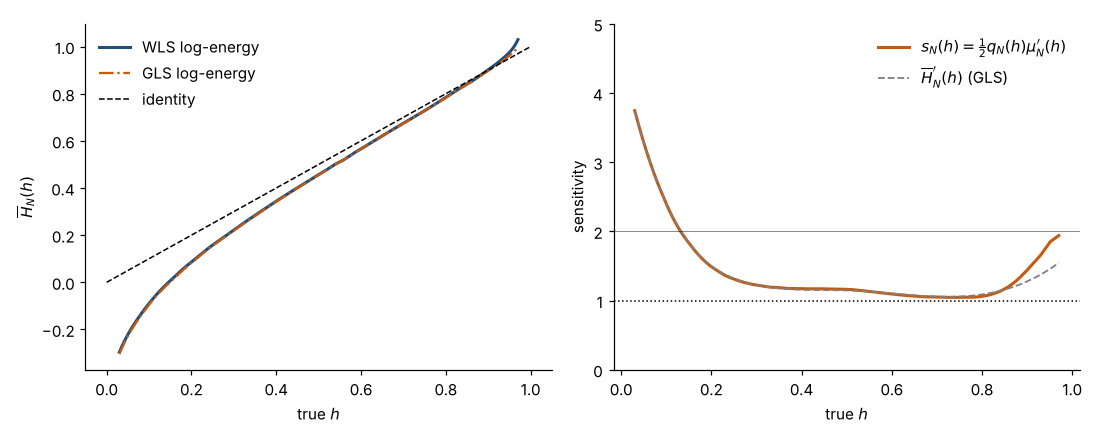}
\caption{Left: expected value $\overline H_N(h)$ of the uncalibrated WLS log-energy estimator as a function of the true $h$ (the binding function), from the calibration bank for $N=1024$, db2, levels $2$--$7$; the dashed line is the identity. Right: the sensitivity $s_N(h)=\frac12q_N(h)\bmu_N'(h)$ (solid), which governs the root expansion of Theorem~\ref{thm:root} and the contraction of the one-step map, and the derivative $\overline H_N'(h)$ of the GLS curve (dashed), which differs from $s_N$ by $\frac12q_N'(h)\br_N(h)$. The one-step update with step $\frac12$ is locally contractive only where $0<s_N<2$. Derivatives were computed from smoothing splines fitted to the simulated means.}
\label{fig:binding}
\end{figure}

The map $T_N(h;\bY)=h+\frac12q_N(h)\{\bY-\bmu_N(h)\}$ suggests the one-step estimator $T_N(\widehat H_1;\bY)$ from a preliminary $\widehat H_1$. Write $\bY=\bmu_N(H)+\beps_N$ and $\widehat H_1=H+\delta$. To first order (Proposition~\ref{prop:onestep} gives the precise statement),
$$
T_N(\widehat H_1;\bY)-H\approx\frac12q_N(H)\beps_N+\kappa_N(H)\,\delta,
\qquad
\kappa_N(H)=1-s_N(H).
$$
When the mean diagram has the ideal slope, $s_N=1$ and $\kappa_N=0$, so one step removes the error of the preliminary estimator to first order. When $\br_N'(H)$ is not small, the factor $\frac12$ is the wrong step length. For $N=1024$, $s_N$ lies between about $1.05$ and $1.3$ for $0.3\le H\le0.85$, but it is about $1.5$ at $H=0.2$, $1.8$ at $H=0.15$, $2.3$ at $H=0.1$, and above $3$ for $H\le0.05$, and it rises again as $H\to1$ (Figure~\ref{fig:binding}; these values come from numerically differentiated simulation means and are accurate to about $\pm0.1$). The map $T_N$ is locally contractive only where $0<s_N<2$, that is, for $H$ above about $0.13$ when $N=1024$. Below that, $|\kappa_N|>1$, the one-step estimator overshoots, and fixed-point iteration of $T_N$ diverges. Section~\ref{sec:gauss} shows the consequence: at $H=0.1$, the one-step estimator started from AV has a larger mean squared error than AV. The remedy is to solve $G_N(h;\bY)=0$ directly, which is a one-dimensional root search on a fine grid. Alternatively one may iterate the scoring-type step $h+\frac12q_N(h)\{\bY-\bmu_N(h)\}/s_N(h)$. This is an approximate Newton step: the exact derivative of $G_N$ is $q_N'(h)\{\bY-\bmu_N(h)\}-2s_N(h)$, and the omitted first term has expectation zero at $h=H$.

\subsection{Asymptotic distribution}

\begin{assumption}[Logscale central limit theorem]
\label{ass:clt}
For fixed $\mathcal J$ there are $a_N\downarrow0$ and a positive definite $V(H)$ with $a_N^{-1/2}\{\bY-\bmu_N(H)\}\Rightarrow N_K(0,V(H))$ and $\bSigma_N(H)=a_NV(H)+o(a_N)$.
\end{assumption}

\begin{assumption}[Smooth calibration]
\label{ass:smooth}
Let $\mathcal H\subset(0,1)$ be compact with $H$ in its interior. For all large $N$, the following three conditions hold. (i) The functions $\bmu_N$ and $\bSigma_N$ are continuously differentiable on $\mathcal H$, $a_N^{-1}\bSigma_N(h)\to V(h)$ uniformly on $\mathcal H$ with $V$ continuous and uniformly positive definite, and $\sup_N\sup_{h\in\mathcal H}\{\|\bmu_N'(h)\|+\|q_N'(h)\|\}<\infty$. (ii) Uniform identification: for every $\epsilon>0$,
$$\liminf_{N\to\infty}\ \inf_{h\in\mathcal H,\,|h-H|\ge\epsilon}\left|q_N(h)\{\bmu_N(H)-\bmu_N(h)\}\right|>0 .$$
(iii) Local regularity: $\liminf_N s_N(H)>0$, and for every sequence $\epsilon_N\to0$,
$$\sup_{|h-H|\le\epsilon_N}\left|q_N(h)\bmu_N'(h)-q_N(H)\bmu_N'(H)\right|\to0 .$$

\end{assumption}

\begin{theorem}[First-order representation of the calibrated estimator]
\label{thm:root}
Under Assumptions~\ref{ass:clt}--\ref{ass:smooth}, with probability tending to one $G_N(\cdot;\bY)$ has exactly one root $\HCAL$ in $\mathcal H$, $\HCAL\to H$ in probability, and
$$
\HCAL-H=\frac{q_N(H)\{\bY-\bmu_N(H)\}}{2s_N(H)}+o_p(a_N^{1/2}).
$$
Consequently $a_N^{-1/2}(\HCAL-H)\Rightarrow N\left(0,\mathcal V_H/s_\infty(H)^2\right)$ with $\mathcal V_H=\frac14e_2^\top\{X^\top V(H)^{-1}X\}^{-1}e_2$ whenever $s_N(H)\to s_\infty(H)$. When $s_N(H)\to1$, the calibrated estimator and the oracle of Proposition~\ref{prop:oracle} are first-order equivalent.
\end{theorem}

\begin{proposition}[One-step update]
\label{prop:onestep}
Under the assumptions of Theorem~\ref{thm:root}, let the preliminary estimator satisfy $\widehat H_1-H=O_p(a_N^{1/2})$. Then
$$
T_N(\widehat H_1;\bY)-H=\frac12q_N(H)\{\bY-\bmu_N(H)\}+\kappa_N(H)(\widehat H_1-H)+o_p(a_N^{1/2}),
\qquad \kappa_N(H)=1-s_N(H).
$$
The one-step estimator is first-order equivalent to the oracle if and only if $\kappa_N(H)(\widehat H_1-H)=o_p(a_N^{1/2})$, for example when $\kappa_N(H)\to0$. Otherwise its error retains the fraction $\kappa_N(H)$ of the preliminary error.
\end{proposition}

The factor $s_N(H)^{-1}$ explains the very small errors of $\HCAL$ at small $H$ in Section~\ref{sec:gauss}. Where the mean diagram is steep in $H$, a given discrepancy $\bY-\bmu_N(h)$ identifies $h$ more precisely. This precision is obtained from the model: it holds when the data are fBm and the calibration bank is built with the same $N$, wavelet, boundary rule, levels, pilot and floor. Under misspecification, if $\E\bY=\bmu_N(H)+\bm\delta$, Theorem~\ref{thm:root} gives the bias $q_N(H)\bm\delta/\{2s_N(H)\}$ to first order. A steep curve damps a given discrepancy rather than amplifying it, but the discrepancy itself, through its projection on $q_N(H)$, can be large (Section~\ref{sec:robust}).

\subsection{Monte Carlo calibration}

In practice $\bmu_N$ and $\bSigma_N$ are estimated from an independent bank of $M_{\mathrm{cal}}$ unit-scale fBm paths at each point of a grid $\mathcal G\subset(0,1)$, interpolated linearly in $h$, and stabilized by a small ridge $\rho\,\mathrm{tr}(\widehat\bSigma)/K\cdot I_K$. If $M_{\mathrm{cal}}\to\infty$, the grid spacing tends to zero, and $\rho\to0$ at rates that make $\sup_h\|\widehat\bmu(h)-\bmu_N(h)\|=o_p(a_N^{1/2})$ and $\sup_h\|a_N^{-1}\widehat\bSigma(h)-V(h)\|=o_p(1)$, then any consistent root of the interpolated estimating equation has the first-order representation in Theorem~\ref{thm:root}. These rates do not control the derivatives of the interpolated calibration, so uniqueness of the root is not guaranteed; Section~\ref{sec:sources} reports how often multiple crossings occurred. The grid should extend close to $0$ and $1$, because a root at the edge of the grid is a clamped estimate.

The procedure is as follows. One first chooses the wavelet, the boundary rule, the levels $\mathcal J$, the pilot and $\lambda$, and, once for each $N$, simulates the calibration bank and tabulates $\widehat\bmu(h)$ and $\widehat q(h)$ on a fine grid. For an observed path one then computes $\bY$ and evaluates $\widehat G(h)=\widehat q(h)\{\bY-\widehat\bmu(h)\}$ on the fine grid $h_{\min}=0.01,\dots,h_{\max}=0.99$. Since $G_N(h;\bmu_N(H))$ is decreasing in $h$ near $H$, the estimate is the first downcrossing of $\widehat G$, from positive to negative values, interpolated linearly between grid points. If $\widehat G(h_{\min})<0$ the estimate is $h_{\min}$, and if $\widehat G>0$ on the whole grid it is $h_{\max}$. If $\widehat G$ has more than one sign change, the first downcrossing is still returned and the event is recorded. No preliminary estimator and no tuning per realization are needed. For the one-step update, a preliminary value outside $[h_{\min},h_{\max}]$ is projected onto this interval and the update is evaluated at the projected value; the one-step estimate itself is not projected. Section~\ref{sec:sources} reports how often each rule was used.

%======================================================================
\section{Gaussian fBm: calibration and comparison of ordinates}
\label{sec:gauss}
%======================================================================

This section evaluates the calibrated estimator and the LEAF ordinates when the data are exactly Gaussian fBm. After describing the design, we compare calibrated and uncalibrated estimators, compare the calibrated ordinates with one another and with the asymptotic theory of Section~\ref{sec:leaf-theory}, and then separate the sources of the gain over the Abry--Veitch estimator.

\subsection{Design}

Fractional Brownian motion was generated exactly by the circulant embedding method of \citet{DaviesHarte1987} applied to fractional Gaussian noise, followed by cumulative summation. We use this single exact generator throughout. Other exact methods (Cholesky factorization, the Hosking recursion, and the Wood--Chan variant of circulant embedding) produce the same distribution and would change only the Monte Carlo noise, so comparisons across generators measure implementation differences rather than properties of the estimators.

The main design uses $N=1024$, the Daubechies wavelet with two vanishing moments (db2), the discrete wavelet transform with half-point symmetric extension, and the levels $\mathcal J=\{2,\dots,7\}$ with $n_j=258,130,66,34,18,10$ coefficients. For $N=512$ and $N=4096$ the levels are $\{2,\dots,J-3\}$. For each $N$ and each $H\in\{0.1,0.2,\dots,0.9\}$ we generated $M=5000$ test paths. All estimators are applied to the same paths, so differences between estimators are paired, and every entry is reported with its Monte Carlo standard error.

The calibration bank is independent of the test paths. It uses the grid $h=0.01,0.03,\dots,0.99$ with $M_{\mathrm{cal}}=2000$ paths per grid point, a ridge $\rho=10^{-6}$, and linear interpolation of $\widehat\bmu$ and $\widehat\bSigma$ to a grid of step $0.001$ on which $\widehat q$ is tabulated. Each ordinate (log-energy; LEAF with the mean pilot and $\lambda\in\{0.1,0.25,0.5,1,2\}$; LEAF with the median pilot and $\lambda\in\{0.05,0.1,0.25,0.5,1,2\}$; average-log; log-median) has its own bank.

We report AV, both unrestricted and restricted to the calibration range $[0.01,0.99]$; the uncorrected log-energy slope with AV weights (WLS); the one-step update of Section~\ref{sec:cal} started from AV; the mean-only calibrated log-energy estimator with AV weights; and the calibrated estimator $\HCAL$ for each ordinate. By Proposition~\ref{prop:av}, $\HCAL$ for the AV ordinate coincides with $\HCAL$ for log-energy. As a benchmark for the clean Gaussian model we also compute the exact Gaussian maximum likelihood estimator (MLE) of $H$ from the fGn increments, with the scale profiled out. It is evaluated on a grid of step $0.005$ with quadratic refinement, for the first $1000$ paths at $H\in\{0.1,0.2,0.3,0.5,0.7,0.9\}$.

\subsection{Calibration and bias}

\begin{table}[ht]
\centering\footnotesize
\setlength{\tabcolsep}{3pt}
\resizebox{\textwidth}{!}{%
\begin{tabular}{c ccc ccccc}
\toprule
& \multicolumn{3}{c}{Uncalibrated or one-step} & \multicolumn{5}{c}{Calibrated estimator $\HCAL$} \\
\cmidrule(lr){2-4}\cmidrule(lr){5-9}
$H$ & AV & WLS & One-step & Log-energy & \makecell{LEAF mean\\$\lambda=0.5$} & \makecell{LEAF median\\$\lambda=0.1$} & Average-log & Log-median \\
\midrule
0.1 & 34.37 (0.22) & 37.59 (0.23) & 50.45 (0.25) & 0.35 (0.01) & 0.37 (0.01) & 0.55 (0.01) & 0.80 (0.02) & 0.85 (0.02) \\
0.2 & 12.51 (0.13) & 14.39 (0.14) & 10.39 (0.11) & 0.77 (0.02) & 0.80 (0.02) & 1.22 (0.02) & 1.74 (0.04) & 1.83 (0.04) \\
0.3 & 6.42 (0.09) & 7.66 (0.10) & 1.31 (0.03) & 1.17 (0.02) & 1.21 (0.02) & 1.80 (0.04) & 2.54 (0.05) & 2.67 (0.05) \\
0.4 & 4.17 (0.07) & 5.06 (0.08) & 1.32 (0.03) & 1.41 (0.03) & 1.47 (0.03) & 2.21 (0.04) & 3.03 (0.06) & 3.30 (0.07) \\
0.5 & 3.33 (0.06) & 4.01 (0.07) & 1.61 (0.03) & 1.69 (0.03) & 1.73 (0.03) & 2.42 (0.05) & 3.33 (0.07) & 3.47 (0.07) \\
0.6 & 2.86 (0.06) & 3.36 (0.07) & 1.77 (0.04) & 1.80 (0.04) & 1.85 (0.04) & 2.64 (0.05) & 3.57 (0.07) & 3.89 (0.08) \\
0.7 & 2.68 (0.05) & 3.04 (0.06) & 1.98 (0.04) & 2.03 (0.04) & 2.07 (0.04) & 2.89 (0.06) & 3.72 (0.07) & 4.19 (0.08) \\
0.8 & 2.70 (0.05) & 2.91 (0.06) & 1.92 (0.04) & 2.00 (0.04) & 2.04 (0.04) & 2.90 (0.06) & 3.70 (0.07) & 4.25 (0.08) \\
0.9 & 3.47 (0.07) & 3.35 (0.07) & 2.08 (0.05) & 1.83 (0.04) & 1.79 (0.04) & 2.43 (0.05) & 3.03 (0.06) & 3.54 (0.07) \\
\bottomrule
\end{tabular}}
\caption{Gaussian fBm, $N=1024$, db2, levels $2$--$7$, $M=5000$ common paths. Mean squared error $\times10^{3}$ with Monte Carlo standard errors in parentheses. WLS is the log-energy slope with AV weights and no bias correction; One-step is $T_N(\widehat H_{\mathrm{AV}};\bY)$ for the log-energy ordinate.}
\label{tab:e1}
\end{table}

\begin{table}[ht]
\centering\footnotesize
\begin{tabular}{c ccccc}
\toprule
$H$ & AV & WLS & One-step & \makecell{Calibrated\\log-energy} & \makecell{Calibrated LEAF\\median, $\lambda=0.1$}\\
\midrule
0.1 & -180.7 (0.6) & -189.4 (0.6) & 221.0 (0.6) & 1.5 (0.3) & 2.3 (0.3) \\
0.2 & -103.4 (0.6) & -112.1 (0.6) & 98.0 (0.4) & 1.9 (0.4) & 2.9 (0.5) \\
0.3 & -67.1 (0.6) & -75.8 (0.6) & 26.6 (0.3) & 1.7 (0.5) & 2.0 (0.6) \\
0.4 & -46.7 (0.6) & -55.4 (0.6) & 11.0 (0.5) & 1.3 (0.5) & 1.5 (0.7) \\
0.5 & -34.9 (0.6) & -43.6 (0.6) & 4.7 (0.6) & 0.5 (0.6) & 0.4 (0.7) \\
0.6 & -24.4 (0.7) & -33.1 (0.7) & 3.3 (0.6) & 0.7 (0.6) & 0.5 (0.7) \\
0.7 & -16.3 (0.7) & -25.0 (0.7) & 1.3 (0.6) & 0.3 (0.6) & 0.7 (0.8) \\
0.8 & -7.5 (0.7) & -16.2 (0.7) & -0.6 (0.6) & -0.5 (0.6) & -0.7 (0.8) \\
0.9 & 10.8 (0.8) & 2.1 (0.8) & -14.7 (0.6) & -2.1 (0.6) & -1.8 (0.7) \\
\bottomrule
\end{tabular}
\caption{Bias $\times10^{3}$ (Monte Carlo standard error) for the setting of Table~\ref{tab:e1}.}
\label{tab:e1bias}
\end{table}

Tables~\ref{tab:e1} and~\ref{tab:e1bias} give the main results. Every uncalibrated estimator is biased downward for $H\le0.8$, and the bias grows as $H$ decreases, reaching $-0.18$ for AV at $H=0.1$. The AV correction reduces the bias of the WLS slope by a constant $0.009$; Section~\ref{sec:sources} shows that most of the remaining bias is structural. The calibrated estimator has small residual biases, at most $0.003$ in absolute value. Its mean squared error is smaller than that of AV by a factor of $97$ at $H=0.1$, $16$ at $H=0.2$, $5.5$ at $H=0.3$, $2.0$ at $H=0.5$, and between $1.3$ and $1.9$ for $H\ge0.6$. At $H=0.1$, however, $98.5\%$ of the AV estimates are below $0.01$ and the mean AV estimate is $-0.081$. When AV is restricted to the same range $[0.01,0.99]$ as the calibrated estimator, the factor at $H=0.1$ is $23$ (Section~\ref{sec:sources}). The very small errors at small $H$ reflect the factor $s_N(H)^{-1}$ of Theorem~\ref{thm:root} (Figure~\ref{fig:binding}).

The one-step update behaves as predicted in Section~\ref{sec:cal}. Where $s_N(H)$ is close to one ($0.4\le H\le0.8$) it is slightly more accurate than the root, by $2$--$7\%$, because its error retains a small, negatively correlated part of the error of its starting value. At $H=0.1$ the AV starting value lies below the grid for $98.5\%$ of the paths. The update is then evaluated at the projected value $0.01$, where $s_N>3$, and the one-step estimate, which is not projected, overshoots by $+0.22$; it is worse than AV. At $H=0.2$, where $s_N\approx1.5$, the update is evaluated at the AV starting value, which lies near $0.1$ where $s_N>2$, and it overshoots by $+0.10$.

\subsection{LEAF and log-energy under Gaussian fBm}

\begin{table}[ht]
\centering\small
\begin{tabular}{llcccc}
\toprule
Pilot & $\lambda$ & \makecell{Asymptotic\\efficiency} & \makecell{Empirical efficiency,\\mean over $H$} & \makecell{Range\\over $H$} & \makecell{Paired MSE difference\\$\times10^3$ at $H=0.5$}\\
\midrule
mean & 0.1 & 0.827 & 0.886 & [0.845, 0.970] & +0.209 (0.018) \\
mean & 0.25 & 0.913 & 0.947 & [0.920, 1.010] & +0.092 (0.012) \\
mean & 0.5 & 0.959 & 0.977 & [0.959, 1.021] & +0.040 (0.008) \\
mean & 1 & 0.985 & 0.993 & [0.982, 1.019] & +0.013 (0.005) \\
mean & 2 & 0.996 & 0.999 & [0.993, 1.013] & +0.003 (0.002) \\
\midrule
median & 0.05 & 0.591 & 0.671 & [0.622, 0.748] & +0.752 (0.036) \\
median & 0.1 & 0.597 & 0.677 & [0.630, 0.752] & +0.726 (0.036) \\
median & 0.25 & 0.583 & 0.667 & [0.622, 0.742] & +0.751 (0.037) \\
median & 0.5 & 0.557 & 0.646 & [0.601, 0.721] & +0.822 (0.039) \\
median & 1 & 0.519 & 0.614 & [0.570, 0.690] & +0.940 (0.042) \\
median & 2 & 0.478 & 0.577 & [0.533, 0.652] & +1.100 (0.046) \\
\multicolumn{2}{l}{average-log} & 0.405 & 0.502 & [0.442, 0.603] & +1.638 (0.058) \\
\multicolumn{2}{l}{log-median} & 0.368 & 0.459 & [0.417, 0.516] & +1.782 (0.062) \\
\bottomrule
\end{tabular}
\caption{Efficiency of calibrated ordinates relative to calibrated log-energy, $N=1024$, $M=5000$. Asymptotic efficiency is $V_{\mathrm{mean}}(\infty)/V(\lambda)$ from Table~\ref{tab:vlam}; empirical efficiency is the ratio of mean squared errors $\mathrm{MSE}(\mathrm{LE})/\mathrm{MSE}(\mathrm{ordinate})$, averaged over $H=0.1,\dots,0.9$; the last column is the paired difference $\mathrm{MSE}(\mathrm{ordinate})-\mathrm{MSE}(\mathrm{LE})$ with its standard error.}
\label{tab:equiv}
\end{table}

Table~\ref{tab:equiv} compares each calibrated ordinate with calibrated log-energy. The empirical efficiencies follow the asymptotic values of Table~\ref{tab:vlam} closely and in the same order. For the mean pilot, efficiency increases with $\lambda$. Table~\ref{tab:paired} gives the paired differences at each $H$. They are statistically detectable at several values of $H$, because $M=5000$ paired paths resolve differences of a fraction of a percent, but they are small. For $\lambda\ge1$ they are below $2\%$ of the mean squared error of log-energy, and for $\lambda=2$ below $0.7\%$ for $H\le0.8$, in line with the $O(\lambda^{-2})$ expansion of Proposition~\ref{prop:large}. At $H=0.9$ mean-pilot LEAF is slightly more accurate than log-energy, by $1.3$ to $2.1\%$.

\begin{table}[ht]
\centering\small
\begin{tabular}{c ccc}
\toprule
$H$ & $\lambda=0.5$ & $\lambda=1$ & $\lambda=2$\\
\midrule
0.1 & +3.01 (0.54) & +1.05 (0.32) & +0.28 (0.17) \\
0.2 & +4.16 (0.52) & +1.69 (0.31) & +0.58 (0.16) \\
0.3 & +3.14 (0.51) & +1.12 (0.30) & +0.30 (0.16) \\
0.4 & +4.27 (0.49) & +1.83 (0.29) & +0.66 (0.15) \\
0.5 & +2.34 (0.46) & +0.78 (0.28) & +0.19 (0.15) \\
0.6 & +2.40 (0.47) & +0.78 (0.28) & +0.18 (0.15) \\
0.7 & +1.87 (0.47) & +0.51 (0.28) & +0.04 (0.15) \\
0.8 & +1.93 (0.47) & +0.50 (0.28) & -0.00 (0.15) \\
0.9 & -2.10 (0.57) & -1.90 (0.37) & -1.27 (0.23) \\
\bottomrule
\end{tabular}
\caption{Paired difference $\{\mathrm{MSE}(\text{LEAF mean},\lambda)-\mathrm{MSE}(\mathrm{LE})\}/\mathrm{MSE}(\mathrm{LE})$ in percent, with Monte Carlo standard error, for the calibrated estimators; $N=1024$, $M=5000$ common paths.}
\label{tab:paired}
\end{table} For the median pilot, the most efficient floor is $\lambda=0.1$, as predicted by the interior minimum of $V(\lambda)$ at $0.096$, and it is substantially more efficient than either endpoint, the average-log and log-median ordinates. The empirical efficiencies are $0.01$--$0.10$ higher than the asymptotic ones, because the coefficients are dependent and the coarse levels carry few of them, but the ordering is the same.

For clean Gaussian fBm, therefore, the improvement over AV comes from calibration. Among calibrated estimators, mean-pilot LEAF with $\lambda\ge0.5$ and log-energy have similar accuracy, and robust ordinates pay the efficiency costs predicted by Table~\ref{tab:vlam}.

\begin{table}[ht]
\centering\footnotesize
\setlength{\tabcolsep}{3pt}
\resizebox{\textwidth}{!}{%
\begin{tabular}{c cccc cccc}
\toprule
& \multicolumn{4}{c}{$N=512$, levels $2$--$6$} & \multicolumn{4}{c}{$N=4096$, levels $2$--$9$}\\
\cmidrule(lr){2-5}\cmidrule(lr){6-9}
$H$ & AV & LE & \makecell{LEAF mean\\$0.5$} & \makecell{LEAF med.\\$0.1$} & AV & LE & \makecell{LEAF mean\\$0.5$} & \makecell{LEAF med.\\$0.1$}\\
\midrule
0.1 & 44.91 (0.39) & 0.88 (0.02) & 0.92 (0.02) & 1.39 (0.03) & 25.476 (0.085) & 0.075 (0.001) & 0.078 (0.001) & 0.119 (0.002) \\
0.2 & 18.61 (0.25) & 1.87 (0.04) & 1.93 (0.04) & 2.88 (0.06) & 7.986 (0.049) & 0.179 (0.004) & 0.185 (0.004) & 0.277 (0.006) \\
0.3 & 10.76 (0.18) & 2.64 (0.05) & 2.73 (0.05) & 4.11 (0.08) & 3.371 (0.032) & 0.243 (0.005) & 0.250 (0.005) & 0.368 (0.007) \\
0.4 & 7.73 (0.15) & 3.25 (0.07) & 3.36 (0.07) & 5.03 (0.10) & 1.746 (0.023) & 0.307 (0.006) & 0.317 (0.006) & 0.469 (0.009) \\
0.5 & 6.88 (0.14) & 4.05 (0.08) & 4.16 (0.08) & 5.91 (0.11) & 1.067 (0.017) & 0.366 (0.007) & 0.375 (0.007) & 0.534 (0.011) \\
0.6 & 5.99 (0.12) & 4.16 (0.08) & 4.31 (0.09) & 6.36 (0.13) & 0.824 (0.015) & 0.412 (0.008) & 0.423 (0.008) & 0.595 (0.012) \\
0.7 & 5.88 (0.12) & 4.27 (0.09) & 4.35 (0.09) & 6.21 (0.12) & 0.657 (0.013) & 0.436 (0.009) & 0.442 (0.009) & 0.602 (0.012) \\
0.8 & 5.78 (0.12) & 4.10 (0.08) & 4.17 (0.08) & 6.01 (0.12) & 0.604 (0.012) & 0.459 (0.009) & 0.465 (0.009) & 0.648 (0.013) \\
0.9 & 7.38 (0.15) & 3.40 (0.07) & 3.41 (0.08) & 4.83 (0.11) & 0.738 (0.017) & 0.497 (0.011) & 0.472 (0.010) & 0.600 (0.012) \\
\bottomrule
\end{tabular}}
\caption{Mean squared error $\times10^3$ (Monte Carlo standard error) of AV and of calibrated estimators for $N=512$ and $N=4096$; LE is calibrated log-energy. $M=5000$.}
\label{tab:e1N}
\end{table}

Table~\ref{tab:e1N} shows that the same pattern holds for $N=512$ and $N=4096$. The gain of calibration over AV persists at $N=4096$, from a factor of $1.3$ at $H=0.8$ to a factor of $340$ at $H=0.1$, or $108$ when AV is restricted to $[0.01,0.99]$. The gain persists because the structural distortion at the fine levels does not vanish when more coarse levels are added.

\subsection{Where the gain over AV comes from}
\label{sec:sources}

\paragraph{Bias decomposition.} Let $\bar{\bm e}(H)$ be the vector of expected level energies $\E\widehat\sigma_{j,\mathrm{mean}}^2$, estimated from the $5000$ test paths, and let $q_W$ be the WLS slope vector with AV weights. The bias of the WLS slope decomposes exactly as
$$
\underbrace{\tfrac12\{q_W\log_2\bar{\bm e}(H)-1\}-H}_{\text{structural}}
\;+\;
\underbrace{\tfrac12q_W\{\E\bY^{\mathrm{LE}}-\log_2\bar{\bm e}(H)\}}_{\text{logarithmic}} .
$$
The structural part is the bias that would remain if the expected energies were known. The logarithmic part is the effect of taking logarithms of correlated, unequal-variance quadratic forms. The AV correction removes the independent-chi-square prediction of the logarithmic part. Table~\ref{tab:decomp} shows the result. The structural part dominates for $H\le0.6$, from $-0.178$ at $H=0.1$ to $-0.021$ at $H=0.6$. It equals the logarithmic part at $H=0.7$, and it is smaller, or of opposite sign, for $H\ge0.8$. The logarithmic part is nearly constant, between $-0.011$ and $-0.014$. The AV correction removes $0.009$ of it, the amount predicted for independent coefficients with a common variance. It leaves a residual of $-0.002$ to $-0.006$, which reflects both the dependence between coefficients and the unequal variances within a level, notably at the boundary coefficients; Table~\ref{tab:decomp} does not separate these two contributions.

\begin{table}[ht]
\centering\small
\begin{tabular}{c cccccc}
\toprule
$H$ & \makecell{WLS\\bias} & Structural & Logarithmic & \makecell{AV\\correction} & \makecell{Residual\\logarithmic} & \makecell{AV\\bias}\\
\midrule
0.1 & -189.4 & -178.4 & -11.0 & 8.7 & -2.3 & -180.7 \\
0.2 & -112.1 & -101.0 & -11.2 & 8.7 & -2.5 & -103.4 \\
0.3 & -75.8 & -64.4 & -11.4 & 8.7 & -2.7 & -67.1 \\
0.4 & -55.4 & -44.1 & -11.3 & 8.7 & -2.6 & -46.7 \\
0.5 & -43.6 & -32.0 & -11.6 & 8.7 & -2.9 & -34.9 \\
0.6 & -33.1 & -21.3 & -11.8 & 8.7 & -3.1 & -24.4 \\
0.7 & -25.0 & -13.0 & -12.0 & 8.7 & -3.3 & -16.3 \\
0.8 & -16.2 & -3.8 & -12.3 & 8.7 & -3.6 & -7.5 \\
0.9 & 2.1 & 16.4 & -14.4 & 8.7 & -5.7 & 10.8 \\
\bottomrule
\end{tabular}
\caption{Decomposition of the bias ($\times10^3$) of the uncorrected WLS log-energy slope, $N=1024$, db2, levels $2$--$7$, $M=5000$. WLS bias $=$ structural $+$ logarithmic; AV bias $=$ WLS bias $+$ AV correction $=$ structural $+$ residual logarithmic. Monte Carlo standard errors of the biases are about $0.6\times10^{-3}$.}
\label{tab:decomp}
\end{table}

\paragraph{Same bounds, mean versus covariance calibration, and the Gaussian likelihood.} Table~\ref{tab:bsd} reports bias and standard deviation separately for AV, AV restricted to $[0.01,0.99]$, the mean-only calibrated estimator with AV weights, the fully calibrated estimator, and the exact MLE. The restriction matters mainly at $H=0.1$, and slightly at $H=0.2$ and $H=0.9$. At $H=0.1$ it reduces the mean squared error of AV from $34.4\times10^{-3}$ to $8.1\times10^{-3}$, mostly by piling the estimates at the bound, and the calibrated estimator remains $23$ times more accurate. Mean-only calibration achieves nearly all of the improvement; full covariance weighting reduces the mean squared error by a further $2$--$4\%$ for $H\le0.8$ and is slightly worse at $H=0.9$. The exact Gaussian MLE had small empirical bias and a standard deviation $1.6$ to $2.2$ times smaller than the calibrated estimator, a gap that reflects the information discarded by a six-level logscale summary. In return the wavelet estimators cost $O(N)$ operations per path once the bank is built, and they admit the robust ordinates of Section~\ref{sec:robust}. The MLE costs $O(N^2)$ or more per likelihood evaluation and, as Section~\ref{sec:robust} shows, deteriorates sharply under large additive outliers.

\begin{table}[ht]
\centering\footnotesize
\setlength{\tabcolsep}{3pt}
\begin{tabular}{c cc cc cc cc cc}
\toprule
& \multicolumn{2}{c}{AV} & \multicolumn{2}{c}{\makecell{AV restricted\\to $[0.01,0.99]$}} & \multicolumn{2}{c}{\makecell{Mean-only calibrated\\(AV weights)}} & \multicolumn{2}{c}{\makecell{Fully calibrated\\(GLS)}} & \multicolumn{2}{c}{\makecell{Gaussian MLE\\($1000$ paths)}}\\
\cmidrule(lr){2-3}\cmidrule(lr){4-5}\cmidrule(lr){6-7}\cmidrule(lr){8-9}\cmidrule(lr){10-11}
$H$ & Bias & SD & Bias & SD & Bias & SD & Bias & SD & Bias & SD\\
\midrule
0.1 & -180.7 & 41.5 & -89.8 & 2.3 & 1.5 & 19.0 & 1.5 & 18.8 & -0.2 & 11.7 \\
0.2 & -103.4 & 42.6 & -103.0 & 41.5 & 2.0 & 28.2 & 1.9 & 27.7 & -0.8 & 15.1 \\
0.3 & -67.1 & 43.8 & -67.1 & 43.8 & 1.8 & 34.7 & 1.7 & 34.2 & -0.3 & 17.7 \\
0.5 & -34.9 & 45.9 & -34.9 & 45.9 & 0.5 & 41.7 & 0.5 & 41.1 & -0.3 & 19.4 \\
0.7 & -16.3 & 49.2 & -16.3 & 49.2 & 0.3 & 45.8 & 0.3 & 45.1 & -0.1 & 20.1 \\
0.9 & 10.8 & 57.9 & 8.0 & 52.4 & -3.8 & 41.5 & -2.1 & 42.7 & -1.5 & 19.6 \\
\bottomrule
\end{tabular}
\caption{Bias and standard deviation ($\times10^3$), $N=1024$, db2, levels $2$--$7$, $M=5000$ common paths (MLE: the first $1000$). Monte Carlo standard errors are about $\mathrm{SD}/\sqrt M$ for the bias.}
\label{tab:bsd}
\end{table}

\paragraph{Fine scales and boundary coefficients.} To test whether initialization and boundaries explain the bias, Table~\ref{tab:variants} repeats the comparison in two variants. In the first the finest fitted level is omitted (levels $3$--$7$). In the second every coefficient whose support reaches the symmetric boundary extension is removed, which leaves $254,126,62,30,14,6$ coefficients at levels $2$--$7$. Each variant has its own calibration bank.
Omitting level $2$ reduces the absolute bias of AV by $36\%$ at $H=0.1$, from $-0.181$ to $-0.116$, and by $40\%$ at $H=0.5$, from $-0.035$ to $-0.021$, which confirms that the structural distortion is concentrated at the finest scales; it also raises the standard deviation, so that the mean squared error of AV improves only for $H\le0.3$ and that of the calibrated estimator worsens at every $H$. Removing boundary-affected coefficients changes the bias of AV little at small $H$ ($-0.183$ at $H=0.1$) but moves it at large $H$, from $+0.011$ to $-0.008$ at $H=0.9$. Boundary coefficients therefore contribute mainly for large $H$, and the small-$H$ bias is not a boundary effect. In both variants the calibrated estimator is more accurate than restricted AV at every $H$, by large factors at small $H$ and by $10$ to $35\%$ for $H\ge0.7$.

\begin{table}[ht]
\centering\footnotesize
\setlength{\tabcolsep}{3pt}
\resizebox{\textwidth}{!}{%
\begin{tabular}{c ccc cc cc c}
\toprule
& \multicolumn{3}{c}{Levels $2$--$7$, all coefficients} & \multicolumn{2}{c}{Levels $3$--$7$} & \multicolumn{2}{c}{Levels $2$--$7$, interior only} & \\
\cmidrule(lr){2-4}\cmidrule(lr){5-6}\cmidrule(lr){7-8}
$H$ & \makecell{AV\\restricted} & \makecell{Mean-only\\calibrated} & \makecell{Fully\\calibrated} & \makecell{AV\\restricted} & \makecell{Fully\\calibrated} & \makecell{AV\\restricted} & \makecell{Fully\\calibrated} & \makecell{Gaussian\\MLE}\\
\midrule
0.1 & 8.07 (0.00) & 0.36 (0.01) & 0.35 (0.01) & 6.43 (0.04) & 1.07 (0.03) & 8.07 (0.00) & 0.36 (0.01) & 0.137 (0.006) \\
0.2 & 12.34 (0.12) & 0.80 (0.02) & 0.77 (0.02) & 7.57 (0.13) & 2.56 (0.05) & 11.89 (0.12) & 0.79 (0.02) & 0.229 (0.010) \\
0.3 & 6.42 (0.09) & 1.21 (0.02) & 1.17 (0.02) & 5.82 (0.12) & 3.40 (0.07) & 5.95 (0.09) & 1.27 (0.02) & 0.313 (0.013) \\
0.5 & 3.33 (0.06) & 1.74 (0.03) & 1.69 (0.03) & 5.42 (0.11) & 4.41 (0.09) & 3.12 (0.06) & 1.88 (0.04) & 0.374 (0.017) \\
0.7 & 2.68 (0.05) & 2.10 (0.04) & 2.03 (0.04) & 5.76 (0.12) & 4.77 (0.09) & 2.56 (0.05) & 2.22 (0.04) & 0.405 (0.019) \\
0.9 & 2.81 (0.05) & 1.74 (0.03) & 1.83 (0.04) & 4.90 (0.08) & 3.26 (0.07) & 2.54 (0.05) & 2.29 (0.04) & 0.388 (0.017) \\
\bottomrule
\end{tabular}}
\caption{Mean squared error $\times10^3$ (Monte Carlo standard error), $N=1024$, $M=5000$ common paths (MLE: the first $1000$). AV is restricted to $[0.01,0.99]$; each calibrated variant uses its own bank.}
\label{tab:variants}
\end{table}

\paragraph{Behavior of the root search.} For levels $2$--$7$, every one of the $5000$ estimating equations had a single interior downcrossing for $H\le0.7$, for both the mean-only and the fully calibrated estimators. At $H=0.8$ one path had multiple crossings under GLS. At $H=0.9$ the fully calibrated equation returned the upper grid edge for $110$ paths ($2.2\%$) and had multiple crossings for $39$ ($0.8\%$). The mean-only equation returned the edge for $4$ paths and never had multiple crossings, which explains its slightly smaller error at $H=0.9$. The lower edge was never returned. In the interior-coefficient variant, $3\%$ of the paths at $H=0.9$ returned the upper edge. Near $H=1$ a calibration grid extending closer to $1$, or the mean-only version, is therefore preferable.

%======================================================================
\section{Robustness to contamination and non-Gaussianity}
\label{sec:robust}
%======================================================================

This section examines what happens when the data depart from the model used for calibration. We describe the contamination and non-Gaussian scenarios, report the accuracy of the calibrated ordinates, of AV and of the Gaussian likelihood estimator, and identify the conditions under which the robust ordinates help and those under which calibration itself becomes the main source of error.

\subsection{Scenarios}

All estimators in this section are calibrated with the clean Gaussian fBm bank of Section~\ref{sec:gauss}, $N=1024$, and the data are generated from the following models, with $H\in\{0.2,0.5,0.8\}$ and $M=5000$ paths. The scenarios share their underlying Gaussian draws, so comparisons across scenarios as well as across estimators are paired.
In the additive-outlier (AO) scenarios the observed path is $B_H(t)+o_t$, where independently at each time, with probability $p\in\{0.5\%,1\%,2\%\}$, we set $o_t=\pm\kappa$ with random sign and $\kappa\in\{5,10\}$ in units of the standard deviation of the increments, and $o_t=0$ otherwise. In the jump scenario, outliers of the same form are added to the increments rather than to the path, so that each produces a level shift; the added compound-Poisson walk changes the scaling of the observed process, and the target is then the latent fBm exponent. In the ARFIMA scenarios the increments are ARFIMA$(0,d,0)$ with $d=H-\frac12$ \citep{GrangerJoyeux1980,Hosking1981} and Gaussian, $t_5$ or $t_3$ innovations, generated by a truncated moving average with a burn-in of $8192$. Finally, in the non-Gaussian-marginal scenario the increments are $F_{t_3}^{-1}\{\Phi(\xi_t)\}$, with $\xi_t$ fractional Gaussian noise. The transform is monotone with Hermite rank one, so the long-range dependence of $\xi$ is inherited \citep{Taqqu1975}. For $H<\frac12$, however, the transform destroys antipersistence: the lag covariances of $\xi$ sum to zero, but the higher Hermite components contribute positive spectral density at frequency zero, so the transformed increments need not have the low-frequency exponent of $\xi$. For this scenario we therefore call $H$ the latent Gaussian parameter, and the row $H=0.2$ measures recovery of that parameter under misspecification.

\begin{table}[p]
\centering\footnotesize
\setlength{\tabcolsep}{2.5pt}
\resizebox{\textwidth}{!}{%
\begin{tabular}{lc ccccccc}
\toprule
& & & \multicolumn{6}{c}{Calibrated estimator $\HCAL$}\\
\cmidrule(lr){4-9}
Scenario & $H$ & AV & LE & \makecell{LEAF mean\\$\lambda=0.1$} & \makecell{LEAF mean\\$\lambda=0.5$} & \makecell{LEAF median\\$\lambda=0.1$} & Average-log & Log-median\\
\midrule
Clean fBm & 0.2 & 11.19 (0.06) & \textbf{2.77 (0.03)} & 3.02 (0.03) & 2.83 (0.03) & 3.49 (0.04) & 4.17 (0.04) & 4.28 (0.04) \\
 & 0.5 & 5.77 (0.05) & \textbf{4.11 (0.04)} & 4.36 (0.04) & 4.16 (0.04) & 4.92 (0.05) & 5.77 (0.06) & 5.89 (0.06) \\
 & 0.8 & 5.20 (0.05) & \textbf{4.47 (0.04)} & 4.72 (0.05) & 4.51 (0.04) & 5.39 (0.05) & 6.08 (0.06) & 6.52 (0.06) \\
\addlinespace
AO, $p=0.5\%$, $\kappa=10$ & 0.2 & 21.02 (0.09) & 6.76 (0.04) & 4.10 (0.03) & 5.13 (0.04) & \textbf{3.52 (0.04)} & 4.16 (0.04) & 4.29 (0.04) \\
 & 0.5 & 12.96 (0.08) & 9.24 (0.07) & 5.98 (0.06) & 7.18 (0.06) & \textbf{5.22 (0.05)} & 6.01 (0.06) & 6.04 (0.06) \\
 & 0.8 & 11.83 (0.09) & 11.31 (0.08) & 7.08 (0.06) & 8.72 (0.07) & \textbf{5.81 (0.06)} & 6.46 (0.07) & 6.62 (0.07) \\
\addlinespace
AO, $p=1\%$, $\kappa=10$ & 0.2 & 27.00 (0.09) & 9.52 (0.04) & 5.59 (0.04) & 7.34 (0.04) & \textbf{3.63 (0.03)} & 4.27 (0.04) & 4.23 (0.04) \\
 & 0.5 & 18.53 (0.08) & 13.94 (0.07) & 8.51 (0.06) & 10.83 (0.06) & \textbf{5.94 (0.06)} & 6.60 (0.07) & 6.25 (0.06) \\
 & 0.8 & 17.49 (0.09) & 16.96 (0.08) & 10.20 (0.07) & 13.13 (0.08) & \textbf{6.82 (0.07)} & 7.34 (0.07) & 6.98 (0.07) \\
\addlinespace
AO, $p=2\%$, $\kappa=10$ & 0.2 & 34.74 (0.08) & 12.70 (0.03) & 8.10 (0.03) & 10.40 (0.03) & 4.26 (0.04) & 4.79 (0.04) & \textbf{4.22 (0.04)} \\
 & 0.5 & 26.49 (0.08) & 20.41 (0.06) & 13.15 (0.06) & 16.58 (0.06) & 8.02 (0.06) & 8.48 (0.07) & \textbf{7.06 (0.07)} \\
 & 0.8 & 26.15 (0.09) & 25.28 (0.08) & 16.14 (0.07) & 20.45 (0.08) & 9.71 (0.07) & 9.94 (0.08) & \textbf{8.28 (0.08)} \\
\addlinespace
AO, $p=1\%$, $\kappa=5$ & 0.2 & 16.83 (0.07) & 4.57 (0.03) & 3.60 (0.03) & 3.93 (0.03) & \textbf{3.58 (0.03)} & 4.19 (0.04) & 4.26 (0.04) \\
 & 0.5 & 9.34 (0.06) & 6.08 (0.05) & \textbf{5.22 (0.05)} & 5.49 (0.05) & 5.32 (0.05) & 6.09 (0.06) & 6.14 (0.06) \\
 & 0.8 & 7.99 (0.07) & 7.28 (0.06) & 5.97 (0.06) & 6.45 (0.06) & \textbf{5.89 (0.06)} & 6.49 (0.07) & 6.72 (0.07) \\
\addlinespace
Jumps, $p=1\%$, $\kappa=10$ & 0.2 & \textbf{18.12 (0.08)} & 22.04 (0.07) & 23.70 (0.08) & 23.41 (0.07) & 23.22 (0.10) & 23.10 (0.10) & 21.80 (0.11) \\
 & 0.5 & 6.11 (0.06) & \textbf{4.68 (0.05)} & 6.31 (0.05) & 5.14 (0.05) & 8.93 (0.07) & 9.44 (0.07) & 10.16 (0.08) \\
 & 0.8 & 12.99 (0.08) & 11.72 (0.07) & 6.31 (0.06) & 8.48 (0.07) & \textbf{5.26 (0.05)} & 5.96 (0.06) & 6.55 (0.06) \\
\addlinespace
ARFIMA, Gaussian & 0.2 & \textbf{5.44 (0.05)} & 6.12 (0.04) & 6.26 (0.05) & 6.16 (0.04) & 6.60 (0.05) & 7.11 (0.06) & 7.19 (0.06) \\
 & 0.5 & 5.67 (0.05) & \textbf{4.12 (0.04)} & 4.37 (0.04) & 4.17 (0.04) & 5.01 (0.05) & 5.83 (0.06) & 6.13 (0.06) \\
 & 0.8 & 5.66 (0.06) & \textbf{4.81 (0.05)} & 5.00 (0.05) & 4.83 (0.05) & 5.59 (0.06) & 6.31 (0.07) & 6.64 (0.07) \\
\addlinespace
ARFIMA, $t_5$ & 0.2 & \textbf{5.45 (0.05)} & 6.24 (0.04) & 7.09 (0.05) & 6.66 (0.04) & 7.94 (0.05) & 8.37 (0.06) & 8.77 (0.06) \\
 & 0.5 & 5.84 (0.06) & \textbf{4.26 (0.04)} & 4.55 (0.05) & 4.28 (0.04) & 5.39 (0.05) & 6.14 (0.06) & 6.57 (0.06) \\
 & 0.8 & 5.60 (0.05) & 4.83 (0.05) & 4.57 (0.05) & \textbf{4.53 (0.05)} & 5.21 (0.05) & 5.89 (0.06) & 6.37 (0.06) \\
\addlinespace
ARFIMA, $t_3$ & 0.2 & \textbf{5.50 (0.05)} & 6.32 (0.05) & 8.35 (0.05) & 7.37 (0.05) & 10.17 (0.06) & 10.52 (0.07) & 11.24 (0.07) \\
 & 0.5 & 6.42 (0.07) & 5.24 (0.12) & 5.47 (0.06) & \textbf{4.98 (0.07)} & 7.22 (0.06) & 7.89 (0.07) & 8.63 (0.08) \\
 & 0.8 & 6.46 (0.07) & 6.03 (0.07) & \textbf{4.98 (0.05)} & 5.05 (0.06) & 6.35 (0.06) & 6.83 (0.06) & 7.79 (0.07) \\
\addlinespace
fGn $\to t_3$ marginals & 0.2 & \textbf{7.27 (0.16)} & 10.52 (0.16) & 11.32 (0.09) & 10.94 (0.12) & 11.95 (0.07) & 12.44 (0.08) & 12.59 (0.08) \\
 & 0.5 & 6.21 (0.06) & 4.93 (0.07) & 5.47 (0.05) & \textbf{4.88 (0.06)} & 7.32 (0.06) & 8.04 (0.07) & 8.72 (0.08) \\
 & 0.8 & 7.13 (0.09) & 6.54 (0.09) & \textbf{4.87 (0.05)} & 5.28 (0.07) & 5.86 (0.06) & 6.41 (0.06) & 7.20 (0.07) \\
\bottomrule
\end{tabular}}
\caption{Root mean squared error $\times10^{2}$ (Monte Carlo standard error) under contamination and non-Gaussianity, $N=1024$, $M=5000$ common paths. All calibrated estimators use the clean Gaussian fBm bank. AO: additive outliers with probability $p$ and size $\kappa$ increment standard deviations. Bold marks the smallest value in each row.}
\label{tab:e2}
\end{table}

\subsection{Results}

Table~\ref{tab:e2} reports root mean squared errors (RMSE); results for the remaining LEAF floors change smoothly with $\lambda$. Table~\ref{tab:mleao} compares the exact Gaussian MLE with the wavelet estimators on the first $1000$ paths of several scenarios.

\paragraph{Additive outliers.} The clearest case is $H=0.5$ with $1\%$ additive outliers of size $\kappa=10$. Calibrated median-pilot LEAF with $\lambda=0.1$ lowers the RMSE from $0.139$ for calibrated log-energy to $0.059$; AV has $0.185$ and the Gaussian MLE $0.197$. The conditions behind this result are sparse, isolated outliers, a calibration bank from clean fBm, and the floor $\lambda=0.1$. Outliers bend the logscale diagram at the fine levels and lead every mean-based estimator, and the Gaussian likelihood, to underestimate $H$. Without outliers, the corresponding RMSEs are $0.058$ for AV and $0.041$ for calibrated log-energy. Median-pilot LEAF is the most accurate estimator in every cell with $\kappa=10$ and $p\le1\%$. With smaller outliers ($\kappa=5$, $p=1\%$) it is best at $H=0.2$ and $0.8$, while at $H=0.5$ the mean pilot with $\lambda=0.1$ has the smallest RMSE ($0.052$ against $0.053$). Relative to calibrated log-energy, median-pilot LEAF reduces RMSE by a factor of $1.8$ to $2.6$ for $\kappa=10$ and $p\le1\%$, and it is more accurate than the two classical robust ordinates, average-log and log-median, which are its endpoints $\lambda=0$ and $\lambda=\infty$. At $p=2\%$ the log-median ordinate is best, by up to $15\%$ for $H\ge0.5$, so the preferred floor moves upward with the contamination rate. The mean pilot with a small floor captures part of the gain: at $\lambda=0.1$ the RMSE under $1\%$ contamination is $40\%$ below that of log-energy, at an efficiency cost of $11\%$ in mean squared error under the clean model (Table~\ref{tab:equiv}).

\begin{table}[ht]
\centering\footnotesize
\setlength{\tabcolsep}{3pt}
\begin{tabular}{lc ccccc}
\toprule
& & & & \multicolumn{3}{c}{Calibrated}\\
\cmidrule(lr){5-7}
Scenario & $H$ & Gaussian MLE & AV & LE & \makecell{LEAF mean\\$\lambda=0.5$} & \makecell{LEAF median\\$\lambda=0.1$}\\
\midrule
AO, $p=0.5\%$, $\kappa=10$ & 0.2 & 7.26 (0.07) & 21.20 (0.19) & 6.84 (0.09) & 5.18 (0.08) & 3.54 (0.08) \\
 & 0.5 & 14.09 (0.12) & 13.07 (0.18) & 9.32 (0.15) & 7.35 (0.13) & 5.51 (0.12) \\
 & 0.8 & 22.40 (0.16) & 11.98 (0.20) & 11.44 (0.18) & 8.92 (0.16) & 6.10 (0.15) \\
\addlinespace
AO, $p=1\%$, $\kappa=10$ & 0.2 & 10.17 (0.06) & 27.14 (0.20) & 9.57 (0.09) & 7.38 (0.08) & 3.62 (0.08) \\
 & 0.5 & 19.65 (0.10) & 18.73 (0.19) & 14.10 (0.16) & 11.03 (0.15) & 6.24 (0.13) \\
 & 0.8 & 29.02 (0.14) & 17.51 (0.21) & 16.96 (0.19) & 13.19 (0.17) & 7.06 (0.16) \\
\addlinespace
AO, $p=2\%$, $\kappa=10$ & 0.2 & 13.08 (0.04) & 34.83 (0.19) & 12.72 (0.07) & 10.42 (0.08) & 4.23 (0.08) \\
 & 0.5 & 25.46 (0.09) & 26.78 (0.19) & 20.64 (0.14) & 16.81 (0.14) & 8.29 (0.14) \\
 & 0.8 & 35.71 (0.14) & 26.09 (0.20) & 25.20 (0.19) & 20.45 (0.17) & 9.93 (0.17) \\
\addlinespace
AO, $p=1\%$, $\kappa=5$ & 0.2 & 4.80 (0.05) & 16.87 (0.15) & 4.57 (0.08) & 3.94 (0.08) & 3.60 (0.08) \\
 & 0.5 & 9.37 (0.08) & 9.56 (0.15) & 6.28 (0.12) & 5.72 (0.11) & 5.60 (0.12) \\
 & 0.8 & 16.43 (0.12) & 8.17 (0.16) & 7.46 (0.14) & 6.67 (0.14) & 6.16 (0.15) \\
\addlinespace
ARFIMA, Gaussian & 0.2 & 8.21 (0.05) & 5.32 (0.12) & 6.03 (0.09) & 6.08 (0.09) & 6.57 (0.11) \\
 & 0.5 & 2.00 (0.04) & 5.75 (0.13) & 4.12 (0.09) & 4.15 (0.09) & 4.89 (0.11) \\
 & 0.8 & 5.74 (0.06) & 5.60 (0.12) & 4.76 (0.10) & 4.75 (0.10) & 5.49 (0.11) \\
\addlinespace
ARFIMA, $t_3$ & 0.2 & 8.20 (0.05) & 5.59 (0.12) & 6.43 (0.11) & 7.47 (0.11) & 10.20 (0.13) \\
 & 0.5 & 1.89 (0.04) & 6.06 (0.13) & 4.69 (0.13) & 4.65 (0.11) & 7.14 (0.14) \\
 & 0.8 & 5.70 (0.06) & 6.57 (0.17) & 6.18 (0.16) & 5.09 (0.14) & 6.12 (0.12) \\
\bottomrule
\end{tabular}
\caption{Root mean squared error $\times10^2$ (Monte Carlo standard error) of the exact Gaussian MLE (fGn model) and of wavelet estimators on the first $1000$ paths of selected scenarios, $N=1024$.}
\label{tab:mleao}
\end{table}

\paragraph{The Gaussian likelihood under contamination.} Table~\ref{tab:mleao} shows that the exact MLE, the most accurate estimator on clean fBm (Table~\ref{tab:bsd}), deteriorates sharply under large additive outliers, especially at larger $H$. With $0.5\%$ outliers of size $10$ its RMSE at $H=0.8$ is $0.224$, against $0.061$ for median-pilot LEAF. Under these outliers at $H=0.2$ it remains more accurate than AV, but not than the calibrated wavelet estimators. It is also biased when the increments are ARFIMA rather than fGn with the same $H$ ($H=0.2$ and $0.8$). At $H=0.5$ ARFIMA$(0,0,0)$ is white noise, so the model is correct and the MLE is again the most accurate.

\paragraph{Heavy tails.} For heavy-tailed increments, ARFIMA with $t_3$ innovations or fGn with $t_3$ marginals, the median-pilot ordinates are worse than log-energy at $H=0.5$, by $40$--$50\%$ in RMSE. This is the effect described in Remark~\ref{rem:shape}: fine-level coefficients are heavy-tailed while coarse-level coefficients are close to Gaussian, so the shift $b_j(\lambda)$ varies with $j$ and the slope is biased upward. LEAF with the mean pilot is the best estimator in the $t_3$ scenarios for $H\ge0.5$, with RMSE up to $25\%$ below log-energy at $H=0.8$; its quadratic influence is reduced by $w(\lambda)$, while its mean pilot keeps the shift nearly constant across levels.

\paragraph{Misspecified calibration.} For $H=0.2$ under ARFIMA, the calibrated estimators are worse than AV even with Gaussian innovations. ARFIMA and fGn with the same $H$ differ at fine scales, so the fBm bank removes a distortion that ARFIMA does not share. For log-energy at $H=0.2$, the discrepancy $\bm\delta$ between the simulated ARFIMA mean of $\bY$ and $\bmu_N(0.2)$ has projection $\frac12q_N(0.2)\bm\delta=0.075$. Dividing by $s_N(0.2)\approx1.46$ gives a predicted bias of $0.052$, which matches the observed $0.053$ (Theorem~\ref{thm:root}). The steepness of the curve reduces this bias; the projected discrepancy causes it. The same mechanism explains the row for fGn with $t_3$ marginals at $H=0.2$. There the target is the latent Gaussian parameter, and the transform has changed the low-frequency exponent of the observed increments. Calibration is a model-based correction and should be used with a bank from the assumed model; the model can be broadened, for example to ARFIMA or to fBm plus noise, at the cost of additional calibration parameters.

\paragraph{Jumps.} Jumps in the increments add a compound-Poisson random walk to the path, which changes the scaling at coarse levels. No ordinate recovers the latent exponent $H=0.2$. At $H=0.5$ the added walk has the same scaling exponent as the signal, so the scaling is unchanged and log-energy is best; at $H=0.8$ the jumps act mainly as isolated large coefficients at fine levels, and the robust ordinates do well.

%======================================================================
\section{Discussion}
\label{sec:disc}
%======================================================================

Two separate mechanisms improve wavelet estimation of $H$. The first is calibration. For sampled fBm the finite-sample mean of the logscale diagram departs from the ideal line mainly because the expected level energies themselves deviate at the fine scales; the logarithmic effect addressed by the Abry--Veitch correction is a smaller, nearly constant part. Solving the calibrated estimating equation removes most of the bias in matched fBm simulations, needs no tuning, and is far more accurate than AV, by factors of $1.3$ to $23$ at $N=1024$ when both estimators are restricted to the same range. Nearly all of this gain comes from calibrating the mean; full covariance weighting adds a few percent. Because the expected ordinates can move quickly with $H$, the root must be solved for: the one-step update with step $\frac12$ fails at small $H$. Calibration is model-based. Under misspecification its first-order bias is the projection $q_N(H)\bm\delta/\{2s_N(H)\}$ of the discrepancy in the mean diagram, which can exceed the bias of AV, as the ARFIMA scenarios at $H=0.2$ show.

The second mechanism is the choice of ordinate. For a single level with independent Gaussian coefficients no ordinate improves on log-energy (Proposition~\ref{prop:cr}), and mean-pilot LEAF differs from log-energy by $O(\lambda^{-2})$ (Proposition~\ref{prop:large}). In the clean simulations the calibrated versions have similar accuracy, with LEAF slightly ahead at $H=0.9$. The LEAF family is valuable because it spans the trade-off between efficiency and robustness with a single dimensionless parameter. The floor also makes coefficientwise logarithms considerably more efficient: with the median pilot, the optimal floor $\lambda^\ast\approx0.1$ raises the asymptotic efficiency from $0.41$ (average-log) or $0.37$ (log-median) to $0.60$, and empirically to $0.68$. This ordinate was the most accurate estimator under sparse additive outliers of size ten in up to $1\%$ of the observations.

On clean, exactly Gaussian fBm the exact likelihood is more accurate than any wavelet estimator studied here, by factors of $2.6$ to $5.1$ in mean squared error at $N=1024$. The calibrated wavelet estimators are preferable when robustness to outliers matters, when $N$ is large enough that likelihood evaluation is costly, or when the model is only approximately fBm. In those settings the exact likelihood has no built-in protection.

Our recommendations follow from these results. When fBm is a credible working model and a wavelet estimator is desired, mean-only calibration is a practical default: it delivers almost all of the improvement over AV, requires only the simulated mean, and produced fewer problematic root searches near $H=0.9$. Full covariance weighting adds a few percent where the covariance bank is reliable. When the working model is in doubt, calibration should be used with care, since under misspecification it can be less accurate than AV, as in the ARFIMA scenarios at $H=0.2$. When the data are believed to be Gaussian, LEAF with the mean pilot and $\lambda\in[0.5,2]$ is practically equivalent to log-energy; when heavy tails are suspected, the mean pilot with $\lambda\approx0.1$--$0.25$ is preferable; and when isolated outliers are suspected, the median pilot with $\lambda\approx0.1$ is the best choice among the ordinates studied. Comparing the estimates across these choices is a simple diagnostic for contamination.

Several questions remain. The calibration bank can be extended to other models, such as ARFIMA or fBm observed with noise, or made nonparametric by resampling. The mean of the median-pilot ordinate depends on the shape of the coefficient distribution, so a calibration that adapts to the marginal law would combine robustness to outliers with robustness to heavy tails. Interval estimation follows from Theorem~\ref{thm:root} with variance $V_N(H)/s_N(H)^2$; its coverage under misspecification should be studied. Finally, the coefficientwise construction carries over directly to nondecimated transforms, masked data, and images. There the levelwise counts and covariances are more irregular, and calibration by simulation is especially convenient.

\section*{Acknowledgments}
B. Vidakovic acknowledges the partial support of the H.O. Hartley Chair Foundation and NSF Award 2515246 at Texas A\&M University.

\appendix

\section{Proofs}
\label{app:proofs}

\paragraph{Proof of Theorem~\ref{thm:var}.} Write $d_{j,k}=\sigma_jZ_k$, $k=1,\dots,n=n_j$, and $\widehat\sigma_j^2=\sigma_j^2(1+\eta)$. By scale equivariance, $Y_j(\lambda)-\log_2\sigma_j^2=F_n(\eta)$ with
$$
F_n(t)=\frac1n\sum_{k=1}^n\log_2\left\{Z_k^2+\lambda(1+t)\right\},\qquad
F_n'(t)=\frac1{n\log2}\sum_{k=1}^n\frac{\lambda}{Z_k^2+\lambda(1+t)} .
$$
For the mean pilot $\eta=n^{-1}\sum_k(Z_k^2-1)$. For the median pilot the Bahadur representation of the sample median of $Z_k^2$ gives $\eta=n^{-1}\sum_k\phi(Z_k)+o_p(n^{-1/2})$, with $\phi$ as in Section~\ref{sec:leaf-theory}. In both cases $\eta=O_p(n^{-1/2})$. On $|t|\le\frac12$ the functions $z\mapsto\lambda/\{z^2+\lambda(1+t)\}$ are bounded by $2$ and monotone in $t$, so by the Glivenko--Cantelli theorem $F_n'(t)\to w_t/\log2$ uniformly in $|t|\le\frac12$ almost surely, where $w_t=\E\lambda/\{Z^2+\lambda(1+t)\}$ is continuous with $w_0=w(\lambda)$. By the mean-value theorem, $F_n(\eta)=F_n(0)+F_n'(\tilde t)\eta$ with $|\tilde t|\le|\eta|$, hence
$$
Y_j(\lambda)-\log_2\sigma_j^2-b(\lambda)=\frac1n\sum_{k=1}^n\mathrm{IF}_\lambda(Z_k)+o_p(n^{-1/2}),
$$
where $\mathrm{IF}_\lambda$ is defined in Section~\ref{sec:leaf-theory} and has mean zero and finite variance, since $\E\{\log(Z^2+\lambda)\}^2<\infty$ and $\phi$ has a finite second moment. The central limit theorem and Slutsky's lemma give the result. For the mean pilot, as $\lambda\to\infty$, $\log(z^2+\lambda)-\E\log(Z^2+\lambda)=O(\lambda^{-1})$ and $w(\lambda)\to1$, so $\mathrm{IF}_\lambda(z)\to(z^2-1)/\log2$, the influence function of $\log_2\widehat\sigma^2_{j,\mathrm{mean}}$, whose second moment is $2/(\log2)^2$. At $\lambda=0$ the ordinate is the average of the independent variables $\log_2Z_k^2$, whose variance is $\psi'(1/2)/(\log2)^2=\pi^2/\{2(\log2)^2\}$.

\paragraph{Proof of Proposition~\ref{prop:cr}.} An ordinate with the stated expansion is a regular asymptotically linear estimator of $\log_2\sigma^2+\mathrm{const}$ in the model $\{N(0,\sigma^2)\}$. Writing $\theta=\log\sigma^2$, the score is $\frac12(z^2-1)$ and the Fisher information is $\frac12$, so the asymptotic variance of a regular estimator of $\theta$ is at least $2$, and that of an estimator of $\theta/\log 2$ is at least $2/(\log2)^2$. Equality holds for the maximum likelihood estimator $\log\widehat\sigma^2_{\mathrm{mean}}$.

\paragraph{Proof of Proposition~\ref{prop:if}.} The first two statements are the expansions of Section~\ref{sec:leaf-theory}. For the median pilot, $\phi$ is bounded and $\log(z^2+\lambda)=2\log|z|+O(z^{-2})$.

\paragraph{Proof of Proposition~\ref{prop:av}.} $\E(\bY-\bm g)=\bmu_N-\bm g$ and $\Cov(\bY-\bm g)=\bSigma_N$, so $G_N$ is unchanged.

\paragraph{Proof of Proposition~\ref{prop:oracle}.} $\E_H\{\bY-\bmu_N(H)\}=0$ gives unbiasedness, and $\Var=\frac14q_N\bSigma_Nq_N^\top=\frac14e_2^\top(X^\top\bSigma_N^{-1}X)^{-1}e_2$. Minimality is the Gauss--Markov property of GLS among $q$ with $qX=e_2^\top$. The covariance-only estimator has mean $\frac12\{q_N\bmu_N-1\}=H+\frac12q_N\br_N$.

\paragraph{Proof of Theorem~\ref{thm:root}.} Write $\beps_N=\bY-\bmu_N(H)=O_p(a_N^{1/2})$ (Assumption~\ref{ass:clt}) and $s_\ast=\liminf_Ns_N(H)>0$. Since $q_{c\bSigma}=q_{\bSigma}$ for $c>0$ and $a_N^{-1}\bSigma_N\to V$ uniformly with $V$ uniformly positive definite, $C_q=\sup_N\sup_{h\in\mathcal H}\|q_N(h)\|<\infty$. Let $C_1$ bound $\|\bmu_N'\|$ and $\|q_N'\|$ on $\mathcal H$ (Assumption~\ref{ass:smooth}(i)). Then
$$
\sup_{h\in\mathcal H}\left|G_N(h;\bY)-G_N(h;\bmu_N(H))\right|\le C_q\|\beps_N\|=O_p(a_N^{1/2}).
$$

\emph{Step 1: a fixed neighborhood of local regularity.} The condition in Assumption~\ref{ass:smooth}(iii), required for every sequence $\epsilon_N\to0$, implies the following: for every $\eta>0$ there are $\delta>0$ and $N_0$ such that $\sup_{|h-H|\le\delta}|q_N(h)\bmu_N'(h)-q_N(H)\bmu_N'(H)|\le\eta$ for all $N\ge N_0$. Otherwise there would be $\eta>0$ and indices $N_1<N_2<\cdots$ for which the supremum over $|h-H|\le1/k$ exceeds $\eta$ at $N=N_k$. Setting $\epsilon_N=1/k$ for $N_k\le N<N_{k+1}$ would then give a sequence $\epsilon_N\to0$ that violates (iii). Fix $\eta=s_\ast/2$, and reduce $\delta$ if necessary so that $C_1^2\delta\le s_\ast/2$ and $[H-\delta,H+\delta]\subset\mathcal H$.

\emph{Step 2: monotonicity on the fixed neighborhood.} The derivative of $G_N(\cdot;\bY)$ is
$$
D_N(h)=q_N'(h)\{\beps_N+\bmu_N(H)-\bmu_N(h)\}-q_N(h)\bmu_N'(h).
$$
For $|h-H|\le\delta$ and large $N$, the first term is bounded by $C_1\|\beps_N\|+C_1^2\delta\le s_\ast/2+o_p(1)$, and $q_N(h)\bmu_N'(h)\ge2s_N(H)-s_\ast/2\ge\frac32s_\ast-o(1)$. Hence, with probability tending to one, $D_N(h)\le-s_\ast/2$ on $[H-\delta,H+\delta]$, so $G_N(\cdot;\bY)$ is strictly decreasing there. Since $G_N(H;\bY)=q_N(H)\beps_N=O_p(a_N^{1/2})$, it follows that $G_N(H-\delta;\bY)\ge s_\ast\delta/2+O_p(a_N^{1/2})>0$ and $G_N(H+\delta;\bY)\le-s_\ast\delta/2+O_p(a_N^{1/2})<0$ with probability tending to one. There is therefore exactly one root in $[H-\delta,H+\delta]$.

\emph{Step 3: no roots elsewhere, and consistency.} By Assumption~\ref{ass:smooth}(ii) with $\epsilon=\delta$, $|G_N(h;\bmu_N(H))|\ge c_\delta>0$ for $h\in\mathcal H$ with $|h-H|\ge\delta$ and all large $N$. By the uniform bound above, $G_N(\cdot;\bY)$ then has no zero there with probability tending to one, so the root in Step 2 is the only root in $\mathcal H$. Applying (ii) with any $\epsilon<\delta$ in the same way shows that $P(|\HCAL-H|<\epsilon)\to1$, so $\HCAL$ is consistent.

\emph{Step 4: representation.} By the mean-value theorem,
$$
0=G_N(\HCAL;\bY)=q_N(H)\beps_N+D_N(\bar h)(\HCAL-H)
$$
for some $\bar h$ between $H$ and $\HCAL$. Since $\bar h\to H$ in probability, there is a deterministic sequence $\epsilon_N\to0$ with $P(|\bar h-H|\le\epsilon_N)\to1$. On that event, (iii) gives $q_N(\bar h)\bmu_N'(\bar h)=2s_N(H)+o(1)$, and the first term of $D_N(\bar h)$ is bounded by $C_1\|\beps_N\|+C_1^2|\bar h-H|=o_p(1)$. Thus $D_N(\bar h)=-2s_N(H)+o_p(1)$, which yields the representation. The limit distribution follows from Assumption~\ref{ass:clt}.

\paragraph{Proof of Proposition~\ref{prop:onestep}.} Write $\delta_N=\widehat H_1-H=O_p(a_N^{1/2})$; if the preliminary estimator is projected onto the calibration grid, $\widehat H_1$ denotes the projected value, which satisfies the same bound for interior $H$. The map $T_N(h;\bY)=h+\frac12q_N(h)\{\bY-\bmu_N(h)\}$ is continuously differentiable, with derivative $T_N'(h)=1+\frac12D_N(h)$ and $D_N$ as in the previous proof. By the mean-value theorem,
$$
T_N(\widehat H_1;\bY)-H=T_N(H;\bY)-H+T_N'(\tilde h)\,\delta_N=\frac12q_N(H)\beps_N+T_N'(\tilde h)\,\delta_N
$$
for some $\tilde h$ between $H$ and $\widehat H_1$. Since $\tilde h\to H$ in probability, the argument of Step 4 above gives $D_N(\tilde h)=-2s_N(H)+o_p(1)$, so $T_N'(\tilde h)=\kappa_N(H)+o_p(1)$. Multiplying by $\delta_N=O_p(a_N^{1/2})$ gives $T_N(\widehat H_1;\bY)-H=\frac12q_N(H)\beps_N+\kappa_N(H)\delta_N+o_p(a_N^{1/2})$, which only requires the first-order smoothness of Assumption~\ref{ass:smooth}.

\section{Implementation}
\label{app:impl}

The discrete wavelet transform uses the db2 filters with half-point symmetric extension, identical to MATLAB's \texttt{dwt} with the \texttt{'sym'} mode, and keeps the detail coefficients at odd positions of the filtered extended signal. Fractional Gaussian noise is generated by circulant embedding of its autocovariance of length $2N$ \citep{DaviesHarte1987}. The embedding uses the lag-$N$ autocovariance at its midpoint, and its eigenvalues were nonnegative at every sample size and Hurst exponent used in the simulations.
% PI CHECK: confirm in leaf/fbm.py that the first circulant row is (g_0,...,g_{N-1},g_N,g_{N-1},...,g_1)
% and that no eigenvalue was clipped. With 0 instead of g_N at the midpoint, N=1024, H=0.99 gives a minimum
% eigenvalue of about -0.83 (vs +0.017 with g_N).
The calibration bank stores, for each ordinate and each $N$, the interpolated mean vectors and GLS slope vectors on a grid of step $0.001$ in $[0.01,0.99]$, so evaluating $\widehat G(h)$ on the whole grid for a new path costs one matrix--vector product, and the root is found by the rules of Section~\ref{sec:cal}. Building the bank for $N=1024$ with $M_{\mathrm{cal}}=2000$ takes about a minute on a laptop, and estimation is essentially instantaneous. The installable Python package, with annotated code, tests, and scripts to reproduce all tables and figures using the original random seeds, is available at \url{https://github.com/BraniV/LEAF}.

\printbibliography

@article{AbryVeitch1998,
  author  = {Abry, Patrice and Veitch, Darryl},
  title   = {Wavelet Analysis of Long-Range-Dependent Traffic},
  journal = {IEEE Transactions on Information Theory},
  year    = {1998},
  volume  = {44},
  number  = {1},
  pages   = {2--15},
  doi     = {10.1109/18.650984}
}

@article{Flandrin1992,
  author  = {Flandrin, Patrick},
  title   = {Wavelet Analysis and Synthesis of Fractional Brownian Motion},
  journal = {IEEE Transactions on Information Theory},
  year    = {1992},
  volume  = {38},
  number  = {2},
  pages   = {910--917},
  doi     = {10.1109/18.119751}
}

@article{FengVidakovic2017,
  author  = {Feng, Chen and Vidakovic, Brani},
  title   = {Estimation of the Hurst Exponent Using Trimean Estimators on Nondecimated Wavelet Coefficients},
  journal = {arXiv preprint arXiv:1709.08775},
  year    = {2017},
  eprint  = {1709.08775},
  archivePrefix = {arXiv}
}

@article{VeitchAbry1999,
  author  = {Veitch, Darryl and Abry, Patrice},
  title   = {A Wavelet-Based Joint Estimator of the Parameters of Long-Range Dependence},
  journal = {IEEE Transactions on Information Theory},
  year    = {1999},
  volume  = {45},
  number  = {3},
  pages   = {878--897},
  doi     = {10.1109/18.761330}
}

@book{PercivalWalden2000,
  author    = {Percival, Donald B. and Walden, Andrew T.},
  title     = {Wavelet Methods for Time Series Analysis},
  publisher = {Cambridge University Press},
  address   = {Cambridge},
  year      = {2000}
}

@article{SoltaniSimardBoichu2004,
  author  = {Soltani, Skander and Simard, Patrice and Boichu, Daniel},
  title   = {Estimation of the Self-Similarity Parameter Using the Wavelet Transform},
  journal = {Signal Processing},
  year    = {2004},
  volume  = {84},
  number  = {1},
  pages   = {117--123},
  doi     = {10.1016/j.sigpro.2003.05.002}
}

@incollection{FengMeiVidakovic2018,
  author    = {Feng, Chen and Mei, Ying and Vidakovic, Brani},
  title     = {Mammogram Diagnostics Using Robust Wavelet-Based Estimator of Hurst Exponent},
  booktitle = {New Frontiers of Biostatistics and Bioinformatics},
  editor    = {Zhao, Yichuan and Chen, Ding-Geng},
  publisher = {Springer International Publishing},
  address   = {Cham},
  year      = {2018},
  pages     = {109--140},
  doi       = {10.1007/978-3-319-99389-8_5}
}

@misc{KangVidakovic2017,
  author        = {Kang, Minkyoung and Vidakovic, Brani},
  title         = {MEDL and MEDLA: Methods for Assessment of Scaling by Medians of Log-Squared Nondecimated Wavelet Coefficients},
  year          = {2017},
  eprint        = {1703.04180},
  archivePrefix = {arXiv},
  primaryClass  = {stat.ME}
}

@article{Park2009Robust,
  author = {Juhyun Park and Cheolwoo Park},
  title = {Robust Estimation of the Hurst Parameter and Selection of an Onset Scaling},
  journal = {Statistica Sinica},
  year = {2009},
  volume = {19},
  number = {4},
  pages = {1531--1555}
}

@inproceedings{AbryTaqquPesquet2000,
  author    = {Abry, Patrice and Taqqu, Murad S. and Pesquet-Popescu, B.},
  title     = {Wavelet Based Estimators for Self-Similar $\alpha$-Stable Processes},
  booktitle = {Proceedings of the International Conference on Signal Processing, 16th World Computer Congress},
  year      = {2000},
  pages     = {369--375}
}

@article{StoevPipirasTaqqu2002,
  author  = {Stoev, Stilian and Pipiras, Vladas and Taqqu, Murad S.},
  title   = {Estimation of the Self-Similarity Parameter in Linear Fractional Stable Motion},
  journal = {Signal Processing},
  volume  = {82},
  number  = {12},
  pages   = {1873--1901},
  year    = {2002}
}

@inproceedings{GacheFlandrinGarreau1991,
  author    = {Gache, Nicole and Flandrin, Patrick and Garreau, Dominique},
  title     = {Fractal Dimension Estimators for Fractional Brownian Motions},
  booktitle = {Proceedings of the IEEE International Conference on Acoustics, Speech, and Signal Processing (ICASSP)},
  year      = {1991},
  pages     = {3557--3560}
}

@article{Vimalajeewa2026WaveletHurst,
  author  = {Vimalajeewa, Dixon and Ruggeri, Fabrizio and Vidakovic, Brani},
  title   = {Wavelet-Based Hurst Exponent Estimation},
  journal = {WIREs Computational Statistics},
  year    = {2026},
  volume  = {18},
  number  = {3},
  pages   = {e70072},
  doi     = {10.1002/wics.70072}
}

@article{Zhang2024Hurst,
  author  = {Zhang, Hong-Yan and Feng, Zhi-Qiang and Feng, Si-Yu and Zhou, Yu},
  title   = {Typical Algorithms for Estimating Hurst Exponent of Time Sequence:
             A Data Analyst's Perspective},
  journal = {IEEE Access},
  year    = {2024},
  volume  = {12},
  pages   = {185528--185556},
  doi     = {10.1109/ACCESS.2024.3512542}
}

@article{Averty2025Hurst,
  author  = {Averty, T. and Boudraa, A. O. and Dar{\'e}-Emzivat, D.},
  title   = {Hurst Exponent Estimation Using Natural Visibility Graph
             Embedding in Fisher--Shannon Plane},
  journal = {Signal Processing},
  year    = {2025},
  volume  = {230},
  pages   = {109884},
  doi     = {10.1016/j.sigpro.2024.109884}
}

@article{Ju2023Fractal,
  author  = {Ju, Hanqiu and Honda, Naoki and Yoshimura, Shige H. and
             Kaneko, Miki and Shigematsu, Taiki and Kiyono, Ken},
  title   = {Multidimensional Fractal Scaling Analysis Using Higher Order
             Moving Average Polynomials and Its Fast Algorithm},
  journal = {Signal Processing},
  year    = {2023},
  volume  = {208},
  pages   = {108997},
  doi     = {10.1016/j.sigpro.2023.108997}
}

@inproceedings{Fujimoto2024Scaling,
  author    = {Fujimoto, Yudai and Watanabe, Eiichi and
               Hayano, Junichiro and Kiyono, Ken},
  title     = {Application of Local Scaling Exponent Profile
               to Heart Rate Variability Analysis},
  booktitle = {Proceedings of the 32nd European Signal Processing
               Conference (EUSIPCO)},
  year      = {2024},
  pages     = {726--729},
  doi       = {10.23919/EUSIPCO63174.2024.10715171}
  }

@article{VeitchTaqquAbry2000,
  author  = {Veitch, Darryl and Taqqu, Murad S. and Abry, Patrice},
  title   = {Meaningful {MRA} initialization for discrete time series},
  journal = {Signal Processing},
  volume  = {80},
  number  = {9},
  pages   = {1971--1983},
  year    = {2000}
}

@article{GourierouxMonfortRenault1993,
  author  = {Gouri{\'e}roux, Christian and Monfort, Alain and Renault, Eric},
  title   = {Indirect inference},
  journal = {Journal of Applied Econometrics},
  volume  = {8},
  number  = {S1},
  pages   = {S85--S118},
  year    = {1993},
  doi     = {10.1002/jae.3950080507}
}

@article{Smith1993,
  author  = {Smith, Anthony A.},
  title   = {Estimating nonlinear time-series models using simulated vector autoregressions},
  journal = {Journal of Applied Econometrics},
  volume  = {8},
  number  = {S1},
  pages   = {S63--S84},
  year    = {1993},
  doi     = {10.1002/jae.3950080506}
}

@article{MacKinnonSmith1998,
  author  = {MacKinnon, James G. and Smith, Anthony A.},
  title   = {Approximate bias correction in econometrics},
  journal = {Journal of Econometrics},
  volume  = {85},
  number  = {2},
  pages   = {205--230},
  year    = {1998},
  doi     = {10.1016/S0304-4076(97)00099-7}
}

@article{GourierouxPhillipsYu2010,
  author  = {Gouri{\'e}roux, Christian and Phillips, Peter C. B. and Yu, Jun},
  title   = {Indirect inference for dynamic panel models},
  journal = {Journal of Econometrics},
  volume  = {157},
  number  = {1},
  pages   = {68--77},
  year    = {2010},
  doi     = {10.1016/j.jeconom.2009.10.024}
}

@article{GuerrierEtAl2019,
  author  = {Guerrier, St{\'e}phane and Dupuis-Lozeron, Elise and Ma, Yanyuan and Victoria-Feser, Maria-Pia},
  title   = {Simulation-based bias correction methods for complex models},
  journal = {Journal of the American Statistical Association},
  volume  = {114},
  number  = {525},
  pages   = {146--157},
  year    = {2019}
}

@article{GuerrierEtAl2013,
  author  = {Guerrier, St{\'e}phane and Skaloud, Jan and Stebler, Yannick and Victoria-Feser, Maria-Pia},
  title   = {Wavelet-variance-based estimation for composite stochastic processes},
  journal = {Journal of the American Statistical Association},
  volume  = {108},
  number  = {503},
  pages   = {1021--1030},
  year    = {2013}
}

@article{GuerrierEtAl2022,
  author  = {Guerrier, St{\'e}phane and Molinari, Roberto and Victoria-Feser, Maria-Pia and Xu, Haotian},
  title   = {Robust two-step wavelet-based inference for time series models},
  journal = {Journal of the American Statistical Association},
  volume  = {117},
  number  = {540},
  pages   = {1996--2013},
  year    = {2022}
}

@article{Coeurjolly2008,
  author  = {Coeurjolly, Jean-Fran{\c{c}}ois},
  title   = {Hurst exponent estimation of locally self-similar {G}aussian processes using sample quantiles},
  journal = {The Annals of Statistics},
  volume  = {36},
  number  = {3},
  pages   = {1404--1434},
  year    = {2008},
  doi     = {10.1214/009053607000000587}
}

@article{DaviesHarte1987,
  author  = {Davies, Robert B. and Harte, David S.},
  title   = {Tests for {H}urst effect},
  journal = {Biometrika},
  volume  = {74},
  number  = {1},
  pages   = {95--101},
  year    = {1987},
  doi     = {10.1093/biomet/74.1.95}
}

@article{Hosking1981,
  author  = {Hosking, J. R. M.},
  title   = {Fractional differencing},
  journal = {Biometrika},
  volume  = {68},
  number  = {1},
  pages   = {165--176},
  year    = {1981},
  doi     = {10.1093/biomet/68.1.165}
}

@article{GrangerJoyeux1980,
  author  = {Granger, Clive W. J. and Joyeux, Roselyne},
  title   = {An introduction to long-memory time series models and fractional differencing},
  journal = {Journal of Time Series Analysis},
  volume  = {1},
  number  = {1},
  pages   = {15--29},
  year    = {1980},
  doi     = {10.1111/j.1467-9892.1980.tb00297.x}
}

@article{Taqqu1975,
  author  = {Taqqu, Murad S.},
  title   = {Weak convergence to fractional {B}rownian motion and to the {R}osenblatt process},
  journal = {Zeitschrift f{\"u}r Wahrscheinlichkeitstheorie und verwandte Gebiete},
  volume  = {31},
  number  = {4},
  pages   = {287--302},
  year    = {1975},
  doi     = {10.1007/BF00532868}
}

@article{Hampel1974,
  author  = {Hampel, Frank R.},
  title   = {The influence curve and its role in robust estimation},
  journal = {Journal of the American Statistical Association},
  volume  = {69},
  number  = {346},
  pages   = {383--393},
  year    = {1974},
  doi     = {10.1080/01621459.1974.10482962}
}

@article{Raubitzek2023,
  author  = {Raubitzek, Sebastian and Corpaci, Luiza and Hofer, Rebecca and Mallinger, Kevin},
  title   = {Scaling exponents of time series data: A machine learning approach},
  journal = {Entropy},
  volume  = {25},
  number  = {12},
  pages   = {1671},
  year    = {2023},
  doi     = {10.3390/e25121671}
}

@book{Fuller1996,
  author    = {Fuller, Wayne A.},
  title     = {Introduction to Statistical Time Series},
  edition   = {2nd},
  publisher = {Wiley},
  address   = {New York},
  year      = {1996},
  note      = {Section 9.3: transformation $\log(y^2+c s^2)-c s^2/(y^2+c s^2)$ with $c=0.02$}
}

@incollection{BreidtCarriquiry1996,
  author    = {Breidt, F. Jay and Carriquiry, Alicia L.},
  title     = {Improved quasi-maximum likelihood estimation for stochastic volatility models},
  booktitle = {Modelling and Prediction Honoring {S}eymour {G}eisser},
  editor    = {Lee, Jack C. and Johnson, Wesley O. and Zellner, Arnold},
  publisher = {Springer},
  address   = {New York},
  pages     = {228--247},
  year      = {1996},
  doi       = {10.1007/978-1-4612-2414-3_14}
}

@article{MondalPercival2012,
  author  = {Mondal, Debashis and Percival, Donald B.},
  title   = {M-estimation of wavelet variance},
  journal = {Annals of the Institute of Statistical Mathematics},
  volume  = {64},
  number  = {1},
  pages   = {27--53},
  year    = {2012},
  doi     = {10.1007/s10463-010-0282-9}
}

@article{BartlettKendall1946,
  author  = {Bartlett, M. S. and Kendall, D. G.},
  title   = {The statistical analysis of variance-heterogeneity and the logarithmic transformation},
  journal = {Supplement to the Journal of the Royal Statistical Society},
  volume  = {8},
  number  = {1},
  pages   = {128--138},
  year    = {1946},
  doi     = {10.2307/2983618}
}

@article{KouamoLevyLeducMoulines2013,
  author  = {Kouamo, Olaf and L{\'e}vy-Leduc, C{\'e}line and Moulines, Eric},
  title   = {Central limit theorem for the robust log-regression wavelet estimation of the memory parameter in the {G}aussian semi-parametric context},
  journal = {Bernoulli},
  volume  = {19},
  number  = {1},
  pages   = {172--204},
  year    = {2013},
  doi     = {10.3150/11-BEJ398},
  note    = {check pages}
}

@article{LevyLeducEtAl2011,
  author  = {L{\'e}vy-Leduc, C{\'e}line and Boistard, H{\'e}l{\`e}ne and Moulines, Eric and Taqqu, Murad S. and Reisen, Valderio A.},
  title   = {Robust estimation of the scale and of the autocovariance function of {G}aussian short- and long-range dependent processes},
  journal = {Journal of Time Series Analysis},
  volume  = {32},
  number  = {2},
  pages   = {135--156},
  year    = {2011}
}

@article{AchardCoeurjolly2010,
  author  = {Achard, Sophie and Coeurjolly, Jean-Fran{\c{c}}ois},
  title   = {Discrete variations of the fractional {B}rownian motion in the presence of outliers and an additive noise},
  journal = {Statistics Surveys},
  volume  = {4},
  pages   = {117--147},
  year    = {2010},
  doi     = {10.1214/09-SS059}
}

@article{ShenZhuLee2007,
  author  = {Shen, Haipeng and Zhu, Zhengyuan and Lee, Thomas C. M.},
  title   = {Robust estimation of the self-similarity parameter in network traffic using wavelet transform},
  journal = {Signal Processing},
  volume  = {87},
  number  = {9},
  pages   = {2111--2124},
  year    = {2007},
  doi     = {10.1016/j.sigpro.2007.02.010},
  note    = {check pages}
}

@misc{HamiltonEtAl2022,
  author       = {Hamilton, Erin K. and Jeon, Seonghye and Ram{\'i}rez Cobo, Pepa and Lee, Kichun Sky and Vidakovic, Brani},
  title        = {Robust wavelet-based assessment of scaling with applications},
  howpublished = {arXiv:2201.09320},
  year         = {2022}
}

@article{Wu2020,
  author  = {Wu, Liang},
  title   = {A note on wavelet-based estimator of the {H}urst parameter},
  journal = {Entropy},
  volume  = {22},
  number  = {3},
  pages   = {349},
  year    = {2020},
  doi     = {10.3390/e22030349}
}

@article{MoulinesRoueffTaqqu2008,
  author  = {Moulines, Eric and Roueff, Fran{\c{c}}ois and Taqqu, Murad S.},
  title   = {A wavelet {W}hittle estimator of the memory parameter of a nonstationary {G}aussian time series},
  journal = {The Annals of Statistics},
  volume  = {36},
  number  = {4},
  pages   = {1925--1956},
  year    = {2008},
  doi     = {10.1214/07-AOS527}
}

@article{FayEtAl2009,
  author  = {Fa{\"y}, Gilles and Moulines, Eric and Roueff, Fran{\c{c}}ois and Taqqu, Murad S.},
  title   = {Estimators of long-memory: {F}ourier versus wavelets},
  journal = {Journal of Econometrics},
  volume  = {151},
  number  = {2},
  pages   = {159--177},
  year    = {2009},
  doi     = {10.1016/j.jeconom.2009.03.005}
}

@article{Weron2002,
  author  = {Weron, Rafa{\l}},
  title   = {Estimating long-range dependence: Finite sample properties and confidence intervals},
  journal = {Physica A: Statistical Mechanics and its Applications},
  volume  = {312},
  number  = {1--2},
  pages   = {285--299},
  year    = {2002},
  doi     = {10.1016/S0378-4371(02)00961-5}
}

@article{DurbinEtAl2002,
  author  = {Durbin, Blythe P. and Hardin, Johanna S. and Hawkins, Douglas M. and Rocke, David M.},
  title   = {A variance-stabilizing transformation for gene-expression microarray data},
  journal = {Bioinformatics},
  volume  = {18},
  number  = {suppl\_1},
  pages   = {S105--S110},
  year    = {2002},
  doi     = {10.1093/bioinformatics/18.suppl_1.S105}
}

@article{AhlmannEltzeHuber2023,
  author  = {Ahlmann-Eltze, Constantin and Huber, Wolfgang},
  title   = {Comparison of transformations for single-cell {RNA}-seq data},
  journal = {Nature Methods},
  volume  = {20},
  number  = {5},
  pages   = {665--672},
  year    = {2023},
  doi     = {10.1038/s41592-023-01814-1}
}

@article{RousseeuwCroux1993,
  title     = {Alternatives to the Median Absolute Deviation},
  author    = {Rousseeuw, Peter J. and Croux, Christophe},
  journal   = {Journal of the American Statistical Association},
  volume    = {88},
  number    = {424},
  pages     = {1273--1283},
  year      = {1993},
  publisher = {Taylor \& Francis},
  doi       = {10.1080/01621459.1993.10476408}
}

\end{document}